\documentclass[aps,prl,twocolumn,superscriptaddress,showpacs,longbibliography]{revtex4-1}
\usepackage{amsmath}
\usepackage{graphicx}
\usepackage{amsfonts}
\usepackage{verbatim}
\usepackage{amssymb}
\usepackage{color}
\usepackage{dsfont}
\usepackage{amsmath} 
\usepackage{hyperref}
\hypersetup{colorlinks=true, urlcolor=blue, linkcolor=blue, citecolor=blue}

\newcommand{\su}{\uparrow} 
\newcommand{\sd}{\downarrow} 
\newcommand{\bpm}{\begin{pmatrix}}
\newcommand{\epm}{\end{pmatrix}}

\newcommand{\nn}{\nonumber \\} 
\newcommand{\tp}{^{\intercal}}
\newcommand{\dg}{^{\dagger}}
\newcommand{\hc}{\rm{H.c.}}
\newcommand{\half}{\frac{1}{2}}

\newcommand{\wP}{\widetilde{P}}

\newcommand{\nb}{\overline{n}}
\newcommand{\angstrom}{\mbox{\normalfont\AA}}

\begin{document}


\title{Tunable superconducting coupling of quantum dots 
via Andreev bound states in semiconductor-superconductor nanowires}

\author{Chun-Xiao Liu}\email{Corresponding author: chunxiaoliu62@gmail.com}
\affiliation{Qutech and Kavli Institute of Nanoscience, Delft University of Technology, Delft 2600 GA, The Netherlands.}

\author{Guanzhong Wang}
\affiliation{Qutech and Kavli Institute of Nanoscience, Delft University of Technology, Delft 2600 GA, The Netherlands.}

\author{Tom Dvir}
\affiliation{Qutech and Kavli Institute of Nanoscience, Delft University of Technology, Delft 2600 GA, The Netherlands.}

\author{Michael Wimmer}
\affiliation{Qutech and Kavli Institute of Nanoscience, Delft University of Technology, Delft 2600 GA, The Netherlands.}

\date{\today}

\begin{abstract}
Semiconductor quantum dots have proven to be a useful platform for quantum simulation in the solid state.
However, implementing a superconducting coupling between quantum dots mediated by a Cooper pair has so far suffered from limited tunability and strong suppression.
This has limited applications such as Cooper pair splitting and quantum dot simulation of topological Kitaev chains.
In this work, we propose how to mediate tunable effective couplings via Andreev bound states in a semiconductor-superconductor nanowire connecting two quantum dots.
We show that in this way it is possible to individually control both the coupling mediated by Cooper pairs and by single electrons by changing the properties of the Andreev bound states with easily accessible experimental parameters.
In addition, the problem of coupling suppression is greatly mitigated.
We also propose how to experimentally extract the coupling strengths from resonant current in a three-terminal junction.
Our proposal will enable future experiments that have not been possible so far. 
\end{abstract}

\maketitle


\emph{Introduction.}---Semiconductor quantum dots~\cite{Kouwenhoven2001Few, Wiel2002Electron, Hanson2007Spins} have proven to be a useful platform for quantum simulation in the solid state~\cite{Manousakis2002A, Byrnes2008Quantum, Barthelemy2013Quantum}. 
Controlling dot levels and the transfer of single electrons between dots~\cite{Koppens2006Driven, Martins2016Noise, Reed2016Reduced, Baart2016Single} allows to engineer synthetic Hamiltonians such that the desired functionality is achieved, for example allowing for spin qubit operations~\cite{Loss1998Quantum, DiVincenzo2000Universal, Levy2002Universal, Hayashi2003Coherent, Petta2005Coherent, Mizuta2017Quantum}, or simulating the Fermi-Hubbard model~\cite{Hubbard1963Electron, Yang2011Generic, Hensgens2017Quantum} or exotic magnetism~\cite{Nagaoka1966Ferromagnetism,MATTIS2003EIGENVALUES,Nielsen2007Nanoscale, Oguri2007Kondo, Stecher2010Probing, Dehollain2020Nagaoka}.

Adding a superconducting coupling between quantum dots, i.e., a coupling mediated by a Cooper pair instead of single electrons only, would extend the range of possible Hamiltonians tremendously. 
Examples include operations on Andreev qubits~\cite{Zazunov2003Andreev, Chtchelkatchev2003Andreev, Wendin2007Quantum, Padurariu2010Theoretical, Park2017Andreev}, or implementing exotic superconducting systems such as a topological Kitaev chain~\cite{Sau2012Realizing, Leijnse2012Parity, Fulga2013Adaptive}, which might be utilized to implement topological quantum computation~\cite{Nayak2008Non-Abelian, Alicea2012New, Leijnse2012Introduction, Beenakker2013Search, Elliott2015Colloquium, DasSarma2015Majorana, Ivanov2001NonAbelian, Karzig2017Scalable}.

The basic building block for such a simulation is the coupling between two quantum dots. 
In fact, the coupling between two quantum dots mediated by a Cooper pair is of an intrinsic interest for fundamental physics itself: Used as a Cooper pair splitter, the electrons of the Cooper pair are separated in space while maintaining quantum entanglement~\cite{Recher2001Andreev, Loss2000Probing, Falci2001Correlated, Lesovik2001Electronic, Feinberg2003Andreev, Sauret2004Quantum}, which can be used to perform the Bell inequality test~\cite{Bell1966On, Chtchelkatchev2002Bell, Samuelsson2003Orbital} and has potential applications in quantum teleportation~\cite{Bennett1993Teleporting} and quantum cryptography~\cite{Ekert1992Quantum, Gisin2002Quantum}. 
Despite much experimental progress ~\cite{Beckmann2004Evidence, Russo2005Experimental, Hofstetter2009Cooper, Herrmann2010Carbon, Wei2010Positive, Hofstetter2011Finite, Schindele2012Near, Herrmann2012Spectroscopy, Das2012High, Fulop2014Local, Tan2015Cooper, Fulop2015Magnetic, Borzenets2016High, Bruhat2018Circuit, Tan2021Thermoelectric, Pandey2021Ballistic, Ranni2021Real}, the splitting efficiency of Cooper pair splitters nowadays is still not high enough for the Bell inequality test. 
In addition, a sufficient control of the superconducting coupling between two quantum dots, the prerequisite for quantum simulation, has not been demonstrated experimentally. 
To proceed, a method of controlling superconducting and single electron coupling independently is dearly needed.

In most of the existing proposals and experiments, the couplings between quantum dots are mediated by the quasiparticle continuum of the superconductor~\cite{Recher2001Andreev, Falci2001Correlated, Feinberg2003Andreev, Sau2012Realizing, Leijnse2013Coupling}. 
The disadvantage of this approach is the limited tunability, as the electronic properties of the superconducting continuum cannot be controlled experimentally. 
Moreover, the coupling strengths between dots are strongly suppressed when using metallic superconductors.

\begin{figure}[t]
\begin{center}
\includegraphics[width=\linewidth]{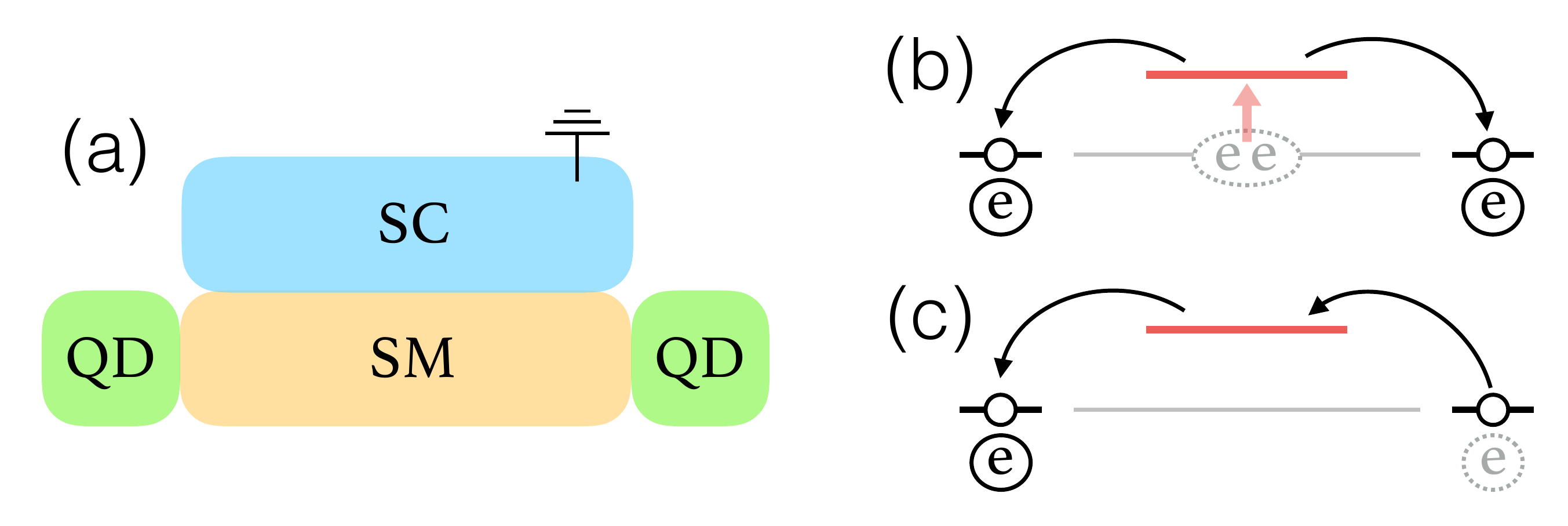}
\caption{
Left: (a) Schematic of the device. 
Two separate quantum dots are connected by a short semiconductor-superconductor hybrid nanowire, which hosts Andreev bound states.
Right: (b) Schematic of cross Andreev reflection and (c) elastic co-tunneling between quantum dots.
The red (black) horizontal line denotes the Andreev bound state (dot level), and the grey line represents the Fermi energy of the superconductor.
}
\label{fig:schematic}
\end{center}
\end{figure}

In this Letter we propose to mediate tunable effective couplings via Andreev bound states in a semiconductor-superconductor nanowire connecting two quantum dots, based on the fact that control over hybrid nanowires has been demonstrated experimentally, e.g., by tuning a nearby electrostatic gates~\cite{de_Moor2018Electric}.
We show that in this way it is possible to individually control both the coupling mediated by Cooper pairs and by single electrons by changing the properties of the Andreev bound states with easily accessible experimental parameters. 
In addition, the problem of coupling suppression is greatly mitigated. 
Finally, we propose how to experimentally extract the coupling strengths from resonant current in a three-terminal junction, allowing for an experimental verification of our theory~\cite{Wang2022Singlet}.

\emph{Model and Hamiltonian.}---The system consists of two quantum dots connected by a semiconductor-superconductor nanowire, as shown in Fig.~\ref{fig:schematic}(a).
The Hamiltonian is
\begin{align}
& H = H_S + H_D + H_{SD}, \nn
&H_S  \approx   E_1 {\gamma}\dg_1 {\gamma}_1 + E_2 {\gamma}\dg_2 {\gamma}_2, \nn
&H_D =  \varepsilon_l d\dg_{l \eta} d_{l \eta} + \varepsilon_r d\dg_{r \sigma} d_{r \sigma}, \nn
&H_{SD} =  - t_{l} c\dg_{x_l \eta}  d_{l \eta}  -t_{r} c\dg_{x_r \sigma}  d_{r \sigma}  + \mathrm{H.c.}.
\label{eq:Ham_DSD}
\end{align}
Here $H_S$ is the Hamiltonian for the hybrid nanowire of length $L=x_r-x_l$.
In the short-wire limit where the level spacing is larger than the superconducting gap, we consider only two normal states closest to the Fermi energy (which form a Kramers' pair in the presence of time-reversal invariance).
With an induced $s$-wave pairing, the normal states are gapped and become two Andreev bound states defined as $\gamma\dg_i = \sum_{x, s=\su, \sd} [ u_i(xs) c\dg_{xs} + v_i(xs) c_{xs}]$, where the wavefunctions and excitation energies are obtained by solving the Bogoliubov-de Gennes equation $h_{\rm{BdG}}(x) (u_i, v_i)\tp = E_{i} (u_i, v_i)\tp$.
$H_D$ describes two quantum dots.
In the limit of strong Zeeman splitting and Coulomb interaction, i.e.,
\begin{align}
\varepsilon_{l,r} < g_{\mathrm{dot}} \mu_B B, U, \quad g_{\mathrm{dot}} \mu_B B < \delta E_{\mathrm{dot}},
\end{align}
each quantum dot accommodates only a single spin-polarized level near Fermi energy~\cite{Hanson2007Spins, supp_tunable}, with the the polarization axes of the two dots being the same and parallel to a globally applied magnetic field. 
Here a large dot level spacing guarantees that adjacent levels are spin-up and -down states from the same orbital.
The spin indices $\eta, \sigma$ in Eq.~\eqref{eq:Ham_DSD} can be either $\su$ or $\sd$, but no summation is taken on them because the dots are in the spin-polarized regime.
$H_{SD}$ describes the spin-conserved electron tunneling between the dot levels and the ends of the hybrid nanowire at $x=x_{l,r}$.

Such setups of two normal dots coupled by a proximitized nanowire segment, i.e. a proximitized central quantum dot, have been studied before experimentally and theoretically in the context of Cooper pair splitting, e.g. in Refs.~\cite{Fulop2015Magnetic, Dominguez2016Quantum}.
In contrast, our focus will be on using the Andreev bound state in the central segment to control the effective coupling of the outer dots.

\emph{Effective couplings between dots.}---In the tunneling limit $t_{l,r} < \Delta$, we can apply a Schrieffer-Wolff transformation to obtain an effective Hamiltonian for the coupled quantum dots.
That is, $H_{\rm{eff}} = H_D + H_{\rm{interdot}}$, with
\begin{align}
H_{\rm{interdot}}  &=  -P H_{SD} \frac{(1-P)}{H_S + H_D} H_{SD} P + O(t^3_{l,r}/\Delta^2) \nn
&= -\Gamma^{\mathrm{CAR}}_{\eta \sigma} d\dg_{l \eta} d\dg_{r \sigma} -\Gamma^{\mathrm{ECT}}_{\eta \sigma} d\dg_{l \eta} d_{r \sigma} + \hc.
\label{eq:H_interdot}
\end{align}
Here $P$ is the projection operator onto the ground state of the uncoupled dot-superconductor system.
$\Gamma^{\mathrm{CAR}}_{\eta \sigma}$ and $\Gamma^{\mathrm{ECT}}_{\eta \sigma}$ are the Andreev bound states-mediated effective couplings between two spin-polarized dot levels, with
\begin{align}
&\Gamma^{\mathrm{CAR}}_{\eta \sigma} = \frac{t_lt_r}{\Delta}  \sum_{m=1,2} \frac{ u_{m }(x_l \eta) v^*_{m }(x_r \sigma) - u_{m}(x_r\sigma) v^*_{m}(x_l \eta) }{E_m/\Delta}, \nn
&\Gamma^{\mathrm{ECT}}_{\eta \sigma}=\frac{t_lt_r}{\Delta}  \sum_{m=1,2} \frac{ u_m(x_l \eta)  u^*_m(x_r \sigma) - v_m(x_r \sigma) v^*_m(x_l \eta) }{E_m/\Delta}.
\label{eq:Gamma_CAR_ECT}
\end{align}
Here $\Gamma^{\mathrm{CAR}}_{\eta \sigma}$ is a superconducting effective coupling, and physically is induced by a coherent crossed Andreev reflection (CAR) process, where an incoming electron with spin-$\sigma$ from the right dot is reflected nonlocally into a hole with spin-$\eta$ in the left dot~[Fig.~\ref{fig:schematic}(b)].
On the other hand, $\Gamma^{\mathrm{ECT}}_{\eta \sigma}$ is a normal effective coupling, and is induced by elastic co-tunneling (ECT), where a single electron hops from the right dot to the left via the Andreev bound states [Fig.~\ref{fig:schematic}(c)].
Equation~\eqref{eq:Gamma_CAR_ECT} is the most general expression.
In what follows, we will define $P^a_{\eta \sigma} = |\Gamma^a_{\eta \sigma} \Delta / (t_l t_r) |^2$ to characterize the coupling strength, and analyze its dependence on the physical parameters of the Andreev bound states.
As we will see, $P^a_{\eta \sigma}$ is proportional to the experimentally measurable current $I^a_{\eta \sigma}$.

\begin{figure}[t]
\begin{center}
\includegraphics[width=\linewidth]{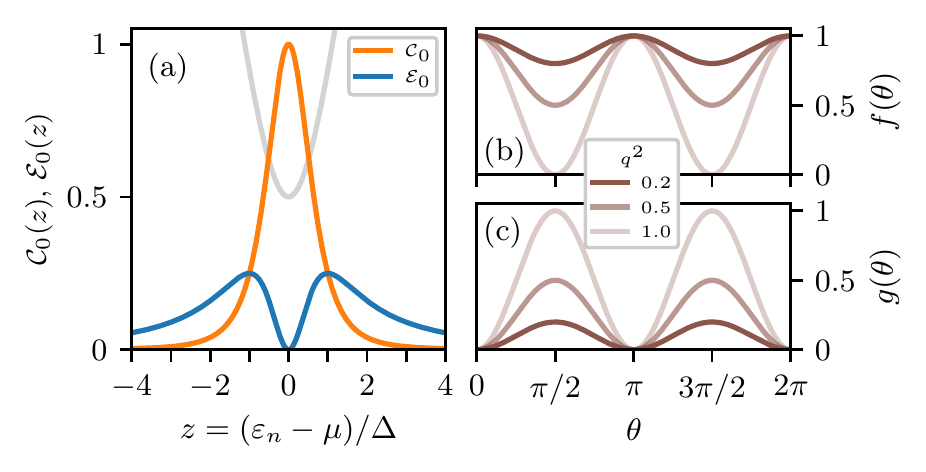}
\caption{
Energy and angle dependence of $P^a$ for a time-reversal invariant hybrid nanowire.
(a) CAR (orange) and ECT (blue) profiles as a function of the normal-state energy $z$.
The grey line denotes the excitation energy $E_m/\Delta =\sqrt{z^2+1}$ of the Andreev bound states (for better visual effect we shift $E_m/\Delta \to E_m/\Delta -1/2$).
Right panels: angle dependence of $P^a$ in favorable (b) and unfavorable (c) channels, with $\theta$ the angle between the spin-orbit field in the hybrid nanowire and the global magnetic field.
Here, $q^2=\sin^2(k_{so}L)$ characterizes the spin-procession through the nanowire due to spin-orbit interaction.
}
\label{fig:TRI} 
\end{center}
\end{figure}

\emph{Energy and angle dependence.}---We first consider a time-reversal invariant hybrid nanowire. 
Physically, this corresponds to a situation where the induced Zeeman splitting in the hybrid segment is negligible compared to the spin-orbit interaction or induced superconducting gap.
The excitation energies of the degenerate Andreev bound states are $E_{1,2}=E_n = \sqrt{\xi^2_n + \Delta^2}$ with $\xi_n = \varepsilon_n-\mu$ being the normal-state energy.
The Bogoliubov-de Gennes wavefunctions are $u_1(x\sigma) = u_0 \psi_n(x\sigma), v_1 = v_0 \psi^*_{\nb}$, and $u_2 = -u_0 \psi_{\nb},v_2 = v_0 \psi^*_n$, where $\psi_{n}, \psi_{\nb}$ are the normal-state wavefunctions, and $u^2_0 = 1 - v^2_0 = 1/2 + \xi_n/2E_n$ are coherence factors.
From Eq.~\eqref{eq:Gamma_CAR_ECT}, we then obtain 
\begin{align}
P^{\rm{CAR}}_{ \eta \sigma} &= \mathcal{C}_0(\xi_n/\Delta) \left|  \psi_n(x_l \eta) \psi_{\nb}(x_r \sigma) - \psi_n(x_r\sigma) \psi_{\nb}(x_l\eta) \right|^2, \nn
P^{\rm{ECT}}_{ \eta \sigma}&=  \mathcal{E}_0(\xi_n/\Delta) | \psi_n(x_l \eta) \psi^*_n(x_r\sigma) + \psi_{\nb}(x_l \eta)  \psi^*_{\nb}(x_r\sigma)  |^2,
\label{eq:P_TRI}
\end{align}
where $\mathcal{C}_0(z)= \left( \frac{2u_0 v_0}{E_n/\Delta} \right)^2=(z^2+1)^{-2}$, $\mathcal{E}_0(z) = \left( \frac{u^2_0 -  v^2_0}{E_n/\Delta} \right)^2 = z^2(z^2+1)^{-2}$ with $z=\xi_n/\Delta$.
Equation~\eqref{eq:P_TRI} shows that $P^a$ has a separable dependence on the energy $\xi_n$ and on the wavefunctions $\psi_{n,\nb }$ of the bound states.
In particular, the energy dependence is universal because it only depends on the coherence factors $u_0$ and $v_0$.
This is a consequence of time reversal symmetry and holds for any hybrid structure, thus not only for one-dimensional wires.
As shown in Fig.~\ref{fig:TRI}(a), $\mathcal{C}_0(z)$ of crossed Andreev reflection has a single peak centered at $z=0$ ($\xi_n = 0$) and decays as $ z^{-4}$ at large $|z|$, while $\mathcal{E}_0(z)$ of elastic co-tunneling has double peaks located at $z=\pm1$, and decays as $ z^{-2}$ at large $|z|$.
Interestingly, $\mathcal{E}_0(z)$ has a dip at $z=0$ due to destructive interference between two virtual paths with a $\pi$-phase shift.
The strikingly different profiles of $\mathcal{C}_0(z)$ and $\mathcal{E}_0(z)$ is the first main finding in this work, which indicates that one can vary the relative CAR and ECT amplitudes by changing the chemical potential of the Andreev bound state.
For the wavefunction part in Eq.~\eqref{eq:P_TRI}, time-reversal invariance, i.e., $\psi_{\nb}(x\sigma)= \mathcal{T}\psi_{n }(x\sigma) = -i\sigma_y \psi^*_{n}(x\sigma)$, gives the following symmetry relations between different dot-spin channels
\begin{align}
P^a_{\su \su} = P^a_{\sd \sd}, \quad P^a_{\su \sd} = P^a_{\sd \su},
\label{eq:sym_relation}
\end{align}
for both CAR and ECT.  
Thus, we will focus on only two spin channels $\su \su$ and $\su \sd$ in the following discussions.

If spin-orbit field is the only spinful field in the hybrid nanowire and has a constant direction, we can find the angle dependence in $P^a$ explicitly.
In the specific case of a one-dimensional Rashba spin-orbit interaction with strength $\alpha_R$~\cite{explanation2022}, the wavefunctions take the form of $\psi_n(x) =  \phi_n(x) e^{- i k_{\mathrm{so}}x \sigma_{\mathrm{so}}} (1, 0)\tp$,
where $\phi_n(x)$ is the eigenfunction in the absence of spin-orbit interaction, $k_{\rm{so}} = m\alpha_R/\hbar^2$ is the spin-orbit wave-vector, and $\sigma_{\mathrm{so}}=\cos \theta \sigma_z + \sin \theta \sigma_x$ is the spin-orbit field which has an angle $\theta$ from the magnetic field.
Here, without loss of generality, we fix the magnetic field (i.e., dot spin axis) along $z$ and rotate the spin-orbit field in the $xz$-plane.
Plugging the wavefunctions into Eq.~\eqref{eq:P_TRI}, we obtain 
\begin{align}
&\wP^{\rm{CAR}}_{\su \su} = \mathcal{C}_0(z) \cdot g(\theta), \quad \wP^{\rm{CAR}}_{\su \sd} = \mathcal{C}_0(z) \cdot f(\theta), \nn
&\wP^{\rm{ECT}}_{\su \su} = \mathcal{E}_0(z) \cdot f(\theta), \quad \wP^{\rm{ECT}}_{\su \sd} = \mathcal{E}_0(z) \cdot g(\theta),
\label{eq:angle}
\end{align}
where $f(\theta) = p^2 + q^2 \cos^2 \theta$ and $g(\theta) = q^2 \sin^2 \theta$. 
Here $p=\cos(k_{\rm{so}} L)$ and $q=\sin(k_{\rm{so}} L)$ characterize the amount of spin precession through the nanowire due to spin-orbit interaction, with $p^2+q^2=1$. 
Note that in Eq.~\eqref{eq:angle}, we have defined a renormalized $\wP^a_{\eta \sigma} =  P^a_{\eta \sigma}/|\phi^2_n(x_l)\phi^2_n(x_r)|$.
The details of the orbital wavefunction $\phi_n(x)$(and thus e.g.~details of the potential landscape or disorder) determine the overall coupling strengths but do not affect the relative CAR and ECT amplitudes. 
As a result, the renormalized $\wP^a$ relies only on the \emph{general} properties of Andreev bound states, i.e., coherence factors $u_0, v_0$, spin-orbit coupling $k_{\mathrm{so}}$ and induced Zeeman spin splitting $E_Z$.
As shown in Figs.~\ref{fig:TRI}(b) and~\ref{fig:TRI}(c), $\wP^a$ has a sinusoidal dependence on the angle $\theta$.
In particular, CAR-$\su \sd$ and ECT-$\su \su$ are more favorable channels with $f(\theta)\geq p^2$.
By contrast, CAR-$\su \su$ and ECT-$\su \sd$ vanish at $\theta=0$ or $\pi$ due to spin conservation.
Hence, in order to have CAR and ECT couplings simultaneously finite in a particular dot spin channel, it is crucial to have a finite spin-orbit field misaligned with the magnetic field.
More surprisingly, although $\wP^a_{\eta \sigma}$ has a strong energy dependence, the ratio of angle-averaged $\wP^a$ in unfavorable and favorable channels depends only on the amount of spin precession, i.e.,
\begin{align}
\frac{ \langle \wP^{\rm{CAR}}_{\su \su} \rangle }{\langle \wP^{\rm{CAR}}_{\su \sd} \rangle } = 
\frac{\langle \wP^{\rm{ECT}}_{\su \sd} \rangle }{\langle \wP^{\rm{ECT}}_{\su \su}  \rangle } = \frac{\sin^2(k_{\rm{so}} L )}{2 - \sin^2(k_{\rm{so}} L )},
\label{eq:ratio}
\end{align}
with $\langle \wP^{a}_{\eta \sigma} \rangle = (2\pi)^{-1} \int^{2\pi}_0 d\theta \wP^{a}_{\eta \sigma} (\theta)$.
This provides a new way to extract the strength of induced spin-orbit coupling in the hybrid nanowire.

\begin{figure}[t]
\begin{center}
\includegraphics[width=\linewidth]{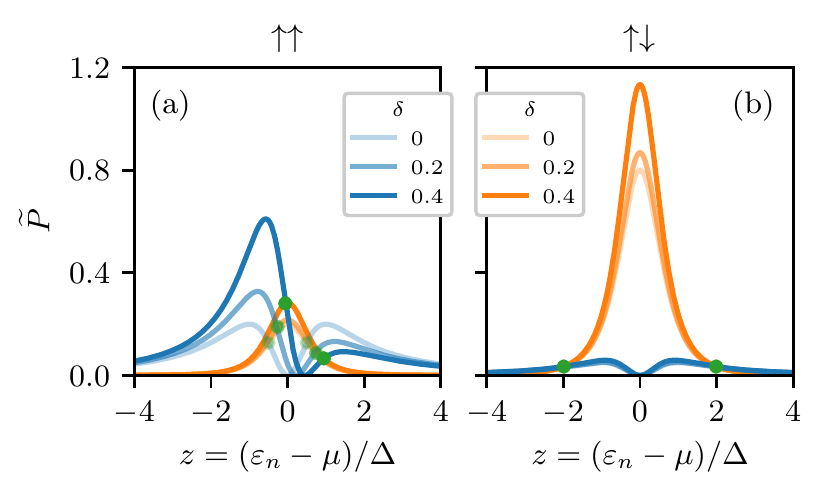}
\caption{Effects of Zeeman spin splitting on CAR (orange) and ECT (blue) profiles in equal-spin (a) and opposite-spin (b) channels.
$\wP^{\rm{CAR}}_{\su \su}$, $\wP^{\rm{CAR}}_{\su \sd}$, and $\wP^{\rm{ECT}}_{\su \sd}$ all increase with $E_Z$, with their profiles remaining symmetric about $z=0$. 
The profile of $\wP^{\rm{ECT}}_{\su \su}$ becomes asymmetric when $E_Z>0$, with one peak being lifted and the other suppressed.
Green dots indicate where $\wP^{\text{CAR}}=\wP^{\text{ECT}}$ for particular values of $\delta$.
Here we choose $q^2=0.2$ and $\theta=\pi/2$, corresponding to the realistic device investigated in~Ref.~\cite{Wang2022Singlet}.}
\label{fig:effect_Ez} 
\end{center}
\end{figure}

\emph{Effect of Zeeman spin splitting.}---We now consider the effect of induced Zeeman splitting in the hybrid segment.
This relaxes the assumption of time-reversal invariance, provides an additional experimentally accessible parameter to tune the profiles of CAR and ECT, and allows for an additional comparison between experiment and theory.
The direction of the Zeeman field is parallel to the spin-polarization axis in dots, i.e., $E_Z\sigma_z$, because we have assumed a globally applied magnetic field in the system. 
However, the magnitude of the Zeeman energy may be different between dots and the hybrid segment because of renormalization effects due to the metallic superconductor \cite{Stanescu2017Proximity}
We also assume weak spin-orbit interaction $k_{so}L \ll 1$ and $E_Z < \Delta$.
Under these assumptions, the energies of the Andreev bound states become $E_{1,2} \approx \sqrt{\xi^2_n + \Delta^2} \pm E_Z$, while the wavefunctions remain the same as those in the time-reversal invariant scenario~\cite{supp_tunable}.
We thus obtain
\begin{align}
&\wP^{\rm{CAR}}_{\su \su}(\delta) = \wP^{\rm{CAR}}_{\sd \sd}(\delta) =  (z^2 + 1 - \delta^2)^2 \cdot q^2 \sin^2 \theta, \nn
&\wP^{\rm{CAR}}_{\su \sd}(\delta) = \wP^{\rm{CAR}}_{\sd \su}(\delta) =(z^2 + 1 - \delta^2)^{-2} \cdot (p^2 + q^2 \cos^2 \theta),  \nn
& \wP^{\mathrm{ECT}}_{\su \su}(\delta) =  \wP^{\mathrm{ECT}}_{\sd \sd}(-\delta)  = \frac{   ( p z - \delta' )^2   + q^2 \cos^2\theta \cdot  (z - \delta)^2 }{(z^2 + 1 - \delta^2)^2}, \nn
& \wP^{\mathrm{ECT}}_{\su \sd}(\delta)=\wP^{\mathrm{ECT}}_{\sd \su} (\delta)  =  \frac{   q^2 z^2 + \delta^2\cos^2  \theta \cdot (1 - p)^2 }{(z^2 + 1 - \delta^2)^2}  \cdot   \sin^2\theta,
\label{eq:EZ_B}
\end{align}
where $\delta=E_Z/\Delta<1$, and $\delta'=\delta( p \cos^2\theta +  \sin^2\theta )$.
As shown in Fig.~\ref{fig:effect_Ez}, $\wP^{\rm{CAR}}_{\su \su}$, $\wP^{\rm{CAR}}_{\su \sd}$, and $\wP^{\rm{ECT}}_{\su \sd}$ all increase with $E_Z$, with their profiles remaining symmetric about $z=0$, while $\wP^{\rm{ECT}}_{\su \su}$ becomes asymmetric, with one peak being lifted and the other suppressed.
In addition, the green dots in Fig.~\ref{fig:effect_Ez} show where $\wP^{\rm{CAR}} = \wP^{\rm{ECT}}$ for particular values of $\delta$, corresponding to the sweet spots in a minimal Kitaev chain.
Such a sweet spot can be found in general because ECT is larger than CAR at large $|z|$ and goes to zero near $z \approx 0$, guaranteeing the crossing of the two curves in most experimentally relevant parameter regimes.

\begin{figure}[t]
\begin{center}
\includegraphics[width=\linewidth]{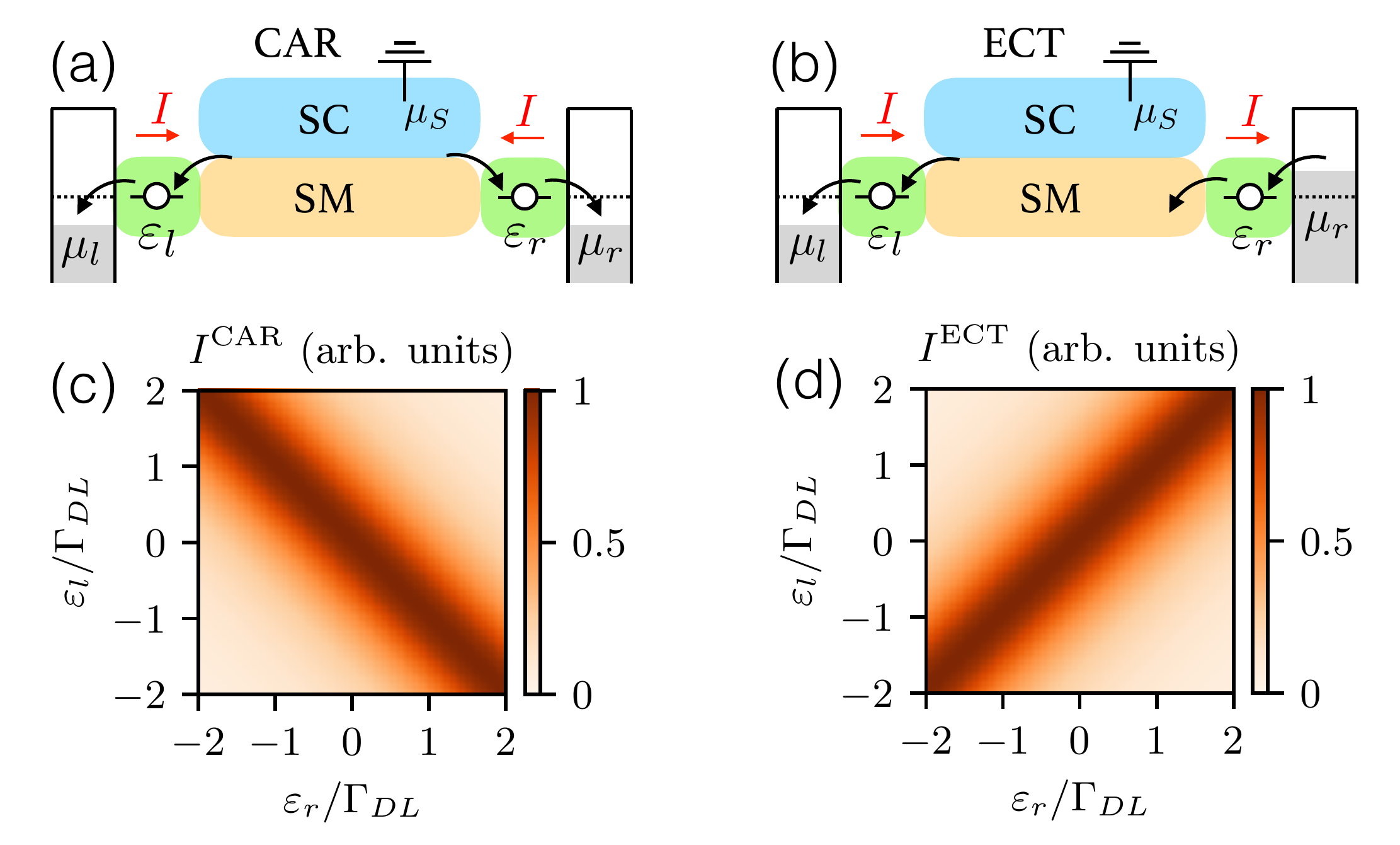}
\caption{
(a) and (b) Schematic for the three-terminal junctions.
(c) and (d) Resonant current in the $(\varepsilon_l, \varepsilon_r)$-plane.
The currents have a Breit-Wigner resonance form, with the broadening width being the dot-lead coupling strength $\Gamma_{DL}$.
CAR and ECT current assumes the maximum value $I^a_{\mathrm{max}}$ when $\varepsilon_l=\pm \varepsilon_r$, respectively.
The strengths of the effective couplings can be extracted by $\Gamma^a = \sqrt{I^a_{\mathrm{max}}  \Gamma_{DL}  \hbar / e  }$
}
\label{fig:experiment}
\end{center}
\end{figure}

\emph{Extracting $\Gamma^a$ experimentally.}---To reach the optimal parameter regime for the desired application, it is necessary to be able to extract the strengths of the effective interdot couplings experimentally.
For this purpose, we propose a three-terminal junction, where two quantum dots are now connected with two external normal electrodes, respectively~[Figs.~\ref{fig:experiment}(a) and~\ref{fig:experiment}(b)].
The strengths of $\Gamma^{\rm{CAR/ECT}}$ can be extracted from resonant current.

Our considerations and calculations follow the same spirit as those in Refs.~\cite{Recher2001Andreev, Loss2000Probing}, which focused on the current due to crossed Andreev reflection in a similar setup.
Compared to the previous works, the differences made in our calculations include:
(1) We now consider Andreev bound states instead of quasiparticle continuum in the superconducting segment. 
(2) Spin-orbit interaction in the hybrid segment breaks spin conservation.
(3) Currents become spin-selective.
(4) We generalize the calculations to elastic co-tunneling scenarios.

The total Hamiltonian for the three-terminal junction, as shown in Fig.~\ref{fig:experiment}, is $H_{\mathrm{tot}}=H + H_L+H_{DL}$. $H$ is the dot-superconductor-dot system introduced by Eq.~\eqref{eq:Ham_DSD}. 
$H_L = \sum_k \left( \varepsilon_{ k} - \mu_{l} \right) a\dg_{l k \eta}a_{l k \eta} + \left( \varepsilon_{k} - \mu_{r} \right) a\dg_{r k \sigma}a_{r k \sigma}$ are the normal leads, which are conventional Fermi liquids with electrons filled up to the Fermi energy $\mu_{l,r}$.
$H_{DL} = \sum_k \left(  - t'_l d\dg_{l \eta}  a_{l k \eta} -t'_r d\dg_{r \sigma}  a_{r k \sigma} \right) + \mathrm{H.c.}$ describes the dot-lead tunneling.
The relevant parameter regime for generating resonant current is~\cite{Recher2001Andreev, Loss2000Probing}
\begin{align}
& \Gamma_{DL}, k_B T < \delta \mu  < \Delta, g_{\mathrm{dot}} \mu_B B, U,\nn
& \varepsilon_{l}, \varepsilon_{r},  \Gamma_{SD} < \Gamma_{DL}.
\label{eq:param}
\end{align}
Here $\delta \mu$ is the applied bias voltage, with $\delta \mu = \mu_S - \mu_{l,r} >0$ for generating CAR current~[Figs.~\ref{fig:experiment}(a)], and $\delta \mu/2 = \mu_r - \mu_S = \mu_S - \mu_l >0$ for ECT~[Figs.~\ref{fig:experiment}(b)].
Bias voltage is smaller than the induced gap $\Delta$, dot charging energy $U$, and dot Zeeman splitting $g_{\mathrm{dot}} \mu_B B$, such that undesired processes such as local Andreev reflection and inelastic co-tunneling would be suppressed, and that the current become spin-selective.
On the other hand, the bias voltage window should be large enough to include the full width of the broadened dot states, i.e., $\delta \mu > \Gamma_{DL} =  \pi \nu (|t'_l|^2 + |t'_r|^2)$ with $\nu$ being the lead density of states.
The dot-lead coupling should be stronger than the superconductor-dot coupling $\Gamma_{DL} >\Gamma_{SD} \approx t_{ln} t_{rn}/\Delta$, such that the quick interdot tunneling process maintains coherence.
Additionally, dot energies need to be tuned close to the superconducting Fermi energy to make dot levels on resonance.
Once all these criteria are met, resonant current will flow between source and drain leads.

The resonant currents are calculated using the rate equation altogether with the $T$-matrix approach~\cite{Recher2001Andreev, Loss2000Probing, ModernQM, supp_tunable}.
When $\mu_S > \mu_{l,r}$, Cooper pairs from the superconducting lead would split into two electrons, which flow to two separate normal leads via dots, respectively, giving the following spin-selective CAR current
\begin{align}
&I^{\mathrm{CAR}}_{\eta \sigma} = \frac{e}{\hbar} \cdot \frac{ \Gamma^2_{DL}   }{( \varepsilon_l + \varepsilon_r)^2 + \Gamma^2_{DL}} \cdot \frac{|\Gamma^{\mathrm{CAR}}_{\eta \sigma}|^2}{ \Gamma_{DL} }, 
\label{eq:I_CAR}
\end{align}
with $\Gamma^{\mathrm{CAR}}_{\eta \sigma}$ being the effective coupling defined in Eq.~\eqref{eq:Gamma_CAR_ECT}.
As shown in Fig.~\ref{fig:experiment}(c), in the $(\varepsilon_l, \varepsilon_r)$-plane CAR current has a Breit-Wigner resonance form with broadening width $\Gamma_{DL} $, and reaches the maximum value along $\varepsilon_l =- \varepsilon_r$ due to energy conservation. 
In exactly the same setup but with a different bias voltage: $\mu_l < \mu_S < \mu_r$, now a single electron flows from one to the other normal lead, giving the spin-selective ECT current
\begin{align}
&I^{\mathrm{ECT}}_{  \eta \sigma } = \frac{e}{\hbar} \cdot \frac{ \Gamma^2_{DL} }{( \varepsilon_l - \varepsilon_r)^2 + \Gamma^2_{DL}} \cdot \frac{|\Gamma^{\mathrm{ECT}}_{\eta \sigma}|^2}{\Gamma_{DL}}, 
\label{eq:I_ECT}
\end{align}
where $\Gamma^{\mathrm{ECT}}_{\eta \sigma}$ is defined in Eq.~\eqref{eq:Gamma_CAR_ECT}.
The ECT current has the same Breit-Wigner form, but now assumes the maximum value when $\varepsilon_l=\varepsilon_r$, as shown in Fig.~\ref{fig:experiment}(d).
Equations~\eqref{eq:I_CAR} and \eqref{eq:I_ECT} indicate that resonant current is proportional to the square of the corresponding interdot coupling strength.
Thus, experimentally one can extract the strengths using the formula $\Gamma^a = \sqrt{I^a_{\mathrm{max}}  \Gamma_{DL}  \hbar / e  }$, where $\Gamma_{DL}$ is read off from the resonance broadening width in gate voltage times the lever arm, and $I^a_{\mathrm{max}}$ is the current value along $\varepsilon_l = -\varepsilon_r$ for CAR and $\varepsilon_l = \varepsilon_r$ for ECT.

\emph{Discussions.}---We have given a proposal for mediating tunable superconducting and normal couplings of quantum dots via Andreev bound states.
This provides an experimentally accessible method for fine-tuning the physical system into the desirable parameter regime.
In particular, the Cooper pair splitting efficiency now can be enhanced by tuning the energy close to $z=0$ in Fig.~\ref{fig:TRI}(a), where the crossed Andreev reflection is strengthened and simultaneously the unwanted elastic co-tunneling processes are strongly suppressed.
On the other hand, a minimal Kitaev chain, which is comprised of two spin-polarized dots, now becomes tunable and can host Majorana zero modes when the superconducting and normal couplings are equal in strength, e.g., where CAR and ECT curves cross each other within the range of $-1 \lesssim z \lesssim 1$ in Fig.~\ref{fig:effect_Ez}(a).
In practice, this tuning protocol can be implemented by controlling the electrostatic gate near the semiconductor-superconductor segment to change the chemical potential therein, eliminating the need of non-collinear magnetic fields~\cite{Leijnse2012Parity}.
This makes our proposal especially appealing, since all the necessary ingredients, i.e., spin-polarized quantum dots~\cite{Hanson2004Semiconductor}, gated hybrid nanowire with spin-orbit interaction~\cite{de_Moor2018Electric, Bommer2019SpinOrbit}, are within reach of existing materials and technologies.
We thus expect that our proposal will enable future experiments that have not been possible so far. 
In fact, in a recent experiment we and our co-workers have already shown a record high Cooper pair splitting efficiency enabled by coupling through Andreev bound states~\cite{Wang2022Singlet}.
Also, a tunable Kitaev chain of two sites has been experimentally realized~\cite{Dvir2022Realization}, providing an exciting platform for studying topological superconductivity and non-Abelian statistics.

\emph{Acknowledgements.}---This work was supported by a subsidy for top consortia for knowledge and innovation (TKl toeslag), by the Dutch Organization for Scientific Research (NWO), by the Foundation for Fundamental Research on Matter (FOM) and by Microsoft Corporation Station Q.

\emph{Author contributions.}---C.-X.L. formulated the project idea with input from G.W. and T.D., and designed the project;
C.-X.L. performed the calculations with input from M.W.;
C.-X.L. and M.W. wrote the manuscript with input from all authors.

\bibliography{references.bib}

\onecolumngrid
\vspace{1cm}
\begin{center}
{\bf\large Supplemental Material for ``Tunable superconducting coupling of quantum dots via Andreev bound states in semiconductor-superconductor nanowires"}
\end{center}
\vspace{0.5cm}

\setcounter{secnumdepth}{3}
\setcounter{equation}{0}
\setcounter{figure}{0}
\renewcommand{\theequation}{S-\arabic{equation}}
\renewcommand{\thefigure}{S\arabic{figure}}
\renewcommand\figurename{Supplementary Figure}
\renewcommand\tablename{Supplementary Table}
\newcommand\Scite[1]{[S\citealp{#1}]}
\newcommand\Scit[1]{S\citealp{#1}}

\makeatletter \renewcommand\@biblabel[1]{[S#1]} \makeatother


\onecolumngrid

\section{Effective Hamiltonian for quantum dots}

\begin{figure}
\begin{center}
\includegraphics[width=0.4\linewidth]{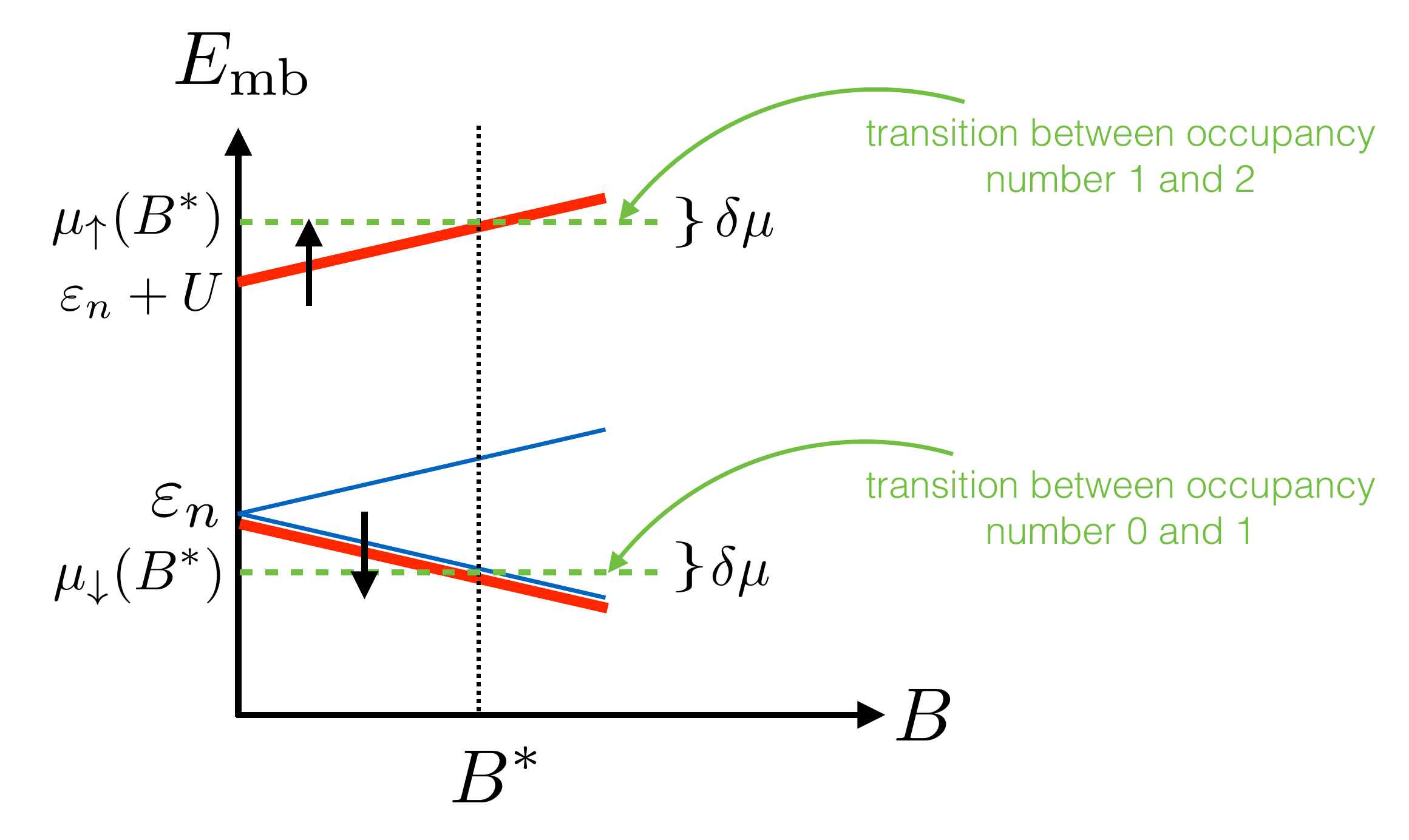}
\caption{Energy diagram for a quantum dot with Coulomb interaction and Zeeman spin splitting.
}
\label{fig:E_mb} 
\end{center}
\end{figure}

The microscopic Hamiltonian for a quantum dot with conventional Coulomb interaction is
\begin{align}
H_{micro} = \left( \varepsilon_n + \half g \mu_B B - \mu \right) \hat{n}_{\su} 
+ \left( \varepsilon_n - \half g \mu_B B - \mu \right) \hat{n}_{\sd}
+ U \hat{n}_{\su} \hat{n}_{\sd},
\end{align}
where $\varepsilon_n$ is the energy of the orbital in the absence of magnetic field, $\half g \mu_B B$ is the Zeeman spin splitting induced by an externally applied magnetic field, giving rise to two spin-polarized states denoted by $n_{\su}$ and $n_{\sd}$, $\mu$ is the chemical potential in the quantum dot, which can be tuned by a nearby eletrostatic gate, and $U$ is the Coulomb interaction.
Here we assume the quantum dot to be in the few-electron regime, and that the level spacing is large, such that there is only a single orbital near the Fermi energy, i.e., 
\begin{align}
g\mu_B B < \delta E_{\mathrm{dot}}.
\end{align}
Figure~\ref{fig:E_mb} shows the many-body energy diagram for such a quantum dot.
Red (blue) lines denote two spin-polarized levels as a function of magnetic field with (without) interaction effect.
When the applied magnetic field is as large as $B=B^*$, and the chemical potential is set $\mu=\mu_{\sd}(B^*) + \delta \mu = \varepsilon_n - \half g \mu_B B^* + \delta \mu $, the energies of the four possible states in the occupation number basis are
\begin{align}
&E_{00} = 0, \quad E_{01} = - \delta \mu, \nn
&E_{10} = g \mu_B B^*, \quad E_{11} = g \mu_B B^*+U -  \delta \mu.
\end{align}
where the bases are defined as $| n_{\su}, n_{\sd} \rangle$.
In the regime of 
\begin{align}
\delta \mu \ll g\mu_B B, U, 
\end{align}
we have
\begin{align}
E_{00} ,E_{01}  \ll E_{10} ,E_{11}.
\end{align}
Thereby $ | 00 \rangle$ and  $| 01 \rangle$ span the low-energy subspace, where the spin-down state can be vacant or occupied while the spin-up state is always vacant.
In the excitation picture, we have the following effective Hamiltonian
\begin{align}
H^{\mathrm{eff}}_{D} \approx - \delta \mu d\dg_{\sd} d_{\sd}, \quad \mathrm{when}~ \mu \approx \mu_{\sd}(B^*) = \varepsilon_n - \half g \mu_B B^*
\end{align}
to describe the transition between $ | 00 \rangle$ and  $| 01 \rangle$ states.
It goes to Eq.~(1) in the main text once we change the notion by $-\delta \mu \to \varepsilon_{l\sd}, \varepsilon_{r\sd}$.
A similar analysis can be applied to the scenario when the Fermi energy is adjusted to a different value $\mu = \mu_{\su}(B^*) = \varepsilon_n + \half g \mu_B B^* + U + \delta \mu$.
The energies for the four possible states are
\begin{align}
&E_{00} = 0, \quad E_{10} = - U -\delta \mu, \nn
&E_{01} = -g \mu_B B^* - U - \delta \mu, \quad E_{11} = -g \mu_B B^* - U - 2 \delta \mu,
\end{align}
and now we have $E_{01}, E_{11} \ll E_{00},E_{10}$.
That is, in the low-energy subspace, spin-down state is always occupied.
Thus the low-energy effective Hamiltonian in the excitation picture is for spin-up state only
\begin{align}
H^{\mathrm{eff}}_{D} \approx - \delta \mu d\dg_{\su} d_{\su}, \quad \mathrm{when}~ \mu \approx \mu_{\su}(B^*) = \varepsilon_n + \half g \mu_B B^* + U.
\end{align}
between $| 01 \rangle$ and  $| 11 \rangle$.
The above discussions justify the dot effective Hamiltonian used in Eq.~(1) in the main text for a single spin-polarized level, with the spin indices being determined by the gate-tunable dot chemical potential, i.e., whether $\mu \approx \mu_{\sd}(B^*)$ or $\mu \approx \mu_{\su}(B^*)$.
The corresponding criteria for this spin-polarized dot level effective Hamiltonian is
\begin{align}
\delta \mu, \Gamma^{\mathrm{CAR/ECT}} \ll g_{dot}\mu_B B^*, U, \delta E_{dot},
\quad g_{dot}\mu_B B^* < \delta E_{dot}
\end{align} 
where $-\delta \mu \to \delta \varepsilon_{l/r}$ in the main text.
In a recent experiment~\cite{Wang2022Singlet}, the values of these parameters extracted from measured data are:
\begin{align}
&\Gamma^{\mathrm{CAR/ECT}} \lesssim 10 \mu eV, \nn
&\delta \mu \approx 0.3 e \times 0.5 mV \times 0.5 = 75 \mu eV,
\end{align}
where $0.3 e$ is the lever arm between gate voltages and the bare quantum dot, $0.5 mV$ is the full range of window where spin-selective resonant current is measured, and multiplication of $0.5$ is to consider the absolute value of $|\delta \mu|$.
\begin{align}
g \mu_B B^* \approx 50 \times 0.06~meV T^{-1} \times 0.1 T = 300 \mu eV.
\end{align}
where $g=50$ is the $g$-factor for bare InSb, $\mu_B$ is the Bohr magneton, and $B=100$ mT is the strength of the applied magnetic field.
The charging energy and level spacing are
\begin{align}
U \approx 2 meV, \quad 2 < \delta E_{dot} < 10 meV,
\end{align}
which are are read from Coulomb diamond diagram.
In addition, level spacing can also be estimated from 
\begin{align}
 \delta E_{dot} \approx \frac{\hbar^2}{2 m  } \cdot \left(  \frac{2\pi}{L}  \right)^2 
 \approx 2.5 meV,
\end{align}
being consistent with the value extracted from Coulomb diamond diagram.
Here the dot length scale is about 200 nm.
As can be seen, the criteria we set for the spin-polarized dot Hamiltonian in Eq.~(1) in the main text is well satisfied by the experimental device in Ref.~\cite{Wang2022Singlet}, justifying a direct comparison between experiment and theory.
\\

\section{General formula for CAR and ECT in a hybrid nanowire}
The Hamiltonian for the hybrid nanowire is
\begin{align}
&H_{\mathrm{hybrid}} = H_{sm} + H_{sc}, \nn
&H_{sm} =  \sum_{\sigma, \sigma' = \su, \sd}  \int^{x_r}_{x_l} dx c\dg_{\sigma}(x) \left( -\frac{\hbar^2}{2m^*} \partial^2_x -i \half[ \alpha_R(x) \partial_x +   \partial_x \alpha_R(x)] \sigma_{so} - \mu(x) + E_Z \sigma_z  \right)_{\sigma \sigma'}  c_{\sigma'}(x), \nn
& H_{sc} = \Delta  \int^{x_r}_{x_l} dx \Big( c\dg_{\su}(x) c\dg_{\sd}(x) +  c_{\sd}(x)c_{\su}(x) \Big).
\end{align}
Here the nanowire is along $x$-axis, $m^*$ is the effective mass of the semiconductor nanowire, $\mu$ is the chemical potential, $E_Z$ is the strength of Zeeman field with its direction $\sigma_z$ parallel to the dot spin axis, $\alpha_R$ is the strength of the spin-orbit coupling.
Without loss of generality, the spin-orbit field has an angle $\theta$ from the Zeeman field, and lies in the $xz$ plane, i.e., $\sigma_{so} = \cos\theta\sigma_z +\sin \theta \sigma_x$.
$\Delta$ is the induced $s$-wave superconducting pairing potential in the nanowire.
To make our discussion as generic as possible, we assume disordered amplitude of chemical potential $\mu(x)$ and spin-orbit coupling $\alpha_R(x)$, while the direction of the spin-orbit field is uniform throughout the nanowire.
In the rest of this section, we will calculate the CAR and ECT couplings in two scenarios, i.e., $E_Z=0$ and $E_Z> 0$.

\subsection{$E_Z=0$}
In the absence of Zeeman field, the normal Hamiltonian of the nanowire becomes
\begin{align}
h_{sm}(\alpha_R>0, E_Z=0) = \frac{\hbar^2}{2m^*}\left[ -i \partial_x + k_{so}(x)\hat{\sigma} \right] \left[ -i \partial_x + k_{so}(x)\hat{\sigma} \right] - \mu(x) - \frac{m \alpha^2_R(x)}{2 \hbar^2}
\end{align}
where $k_{so}(x) = m^* \alpha_R(x)/\hbar^2$ is the local spin-orbit wave-vector.
Its eigenfunction is
\begin{align}
&h_{sm}(\alpha_R>0, E_Z=0)  \psi_n(x) = \xi_n \psi_n(x)  \nn
&\psi_n(x) =  \phi_n(x) e^{- i \beta(x) \sigma_{so}}
\begin{pmatrix}
1 \\
0
\end{pmatrix} =\phi_n(x)
\bpm
\cos( \beta) - i \sin(\beta) \cos (\theta) \\
-i \sin( \beta) \sin (\theta)
\epm
\end{align}
where $\beta(x) = \int^x_0 k_{so}(x')dx'$, and $\phi_n(x) \in \mathbb{R}$ is the eigenfunction in the absence of spin-orbit interaction.
And the time-reversed state is $\psi_{\overline{n}}(x) = -i \sigma_y \psi^*_n(x)$.
From Eq.~(4), we immediately obtain
\begin{align}
&\wP^{\mathrm{CAR}}_{\su \su} = \mathcal{C}_0(z) \cdot  q^2\sin^2(\theta), \nn
&\wP^{\mathrm{CAR}}_{\su \sd} = \mathcal{C}_0(z) \cdot [ p^2+ q^2\cos^2(\theta) ],\nn
&\wP^{\mathrm{ECT}}_{\su \su} = \mathcal{E}_0(z) \cdot [ p^2+ q^2\cos^2(\theta) ], \nn
&\wP^{\mathrm{ECT}}_{\su \sd} = \mathcal{E}_0(z) \cdot q^2\sin^2(\theta),
\end{align}
where $p=  \cos(\overline{k_{so}}L)$ and $q=  \sin(\overline{k_{so}}L)$ characterize the spin procession through the nanowire by spin-orbit interaction, and 
\begin{align}
\overline{k_{so}} = L^{-1} \int^{x_r}_{x_l} dx' k_{so}(x')
\end{align}
is the averaged spin-orbit wave-vector.
Note that in these expressions, we have neglected a prefactor $\phi^2(x_l)\phi^2(x_r)$ common to \emph{all} $P^a_{\eta \sigma} $. 
This \emph{non-universal} local density of states depend on the details of the orbital wavefunction of the bound states, and will determine the overall strength of effective couplings via
\begin{align}
t_l \to t_l \phi(x_l), \quad t_r \to t_r \phi(x_r).
\end{align}
By contrast, what remains in $\wP^a_{\eta \sigma} $ is \emph{universal} and replies only on the coherence factors and spinor part of the Andreev bound states, with the former being determined by chemical potential $\mu$, and the latter by spin-orbit coupling $\alpha_R$ and Zeeman spin splitting $E_Z$.
We now define the angle averaged $P$, i.e.,
\begin{align}
& \langle \wP^{\rm{CAR}}_{ \eta \sigma} \rangle = (2\pi)^{-1} \int^{2\pi}_0  \wP^{\rm{CAR}}_{ \eta \sigma}(\theta) d \theta.
\end{align}
Thus the ratio between $\wP$'s in the unfavorable and favorable channels is
\begin{align}
r = \frac{ \langle \wP^{\rm{CAR}}_{ \su \su} \rangle }{ \langle \wP^{\rm{CAR}}_{ \su \sd} \rangle } =  \frac{ \langle \wP^{\rm{ECT}}_{ \su \sd} \rangle }{ \langle \wP^{\rm{ECT}}_{ \su \su} \rangle } = \frac{ \sin^2( \overline{ k_{so} }  L  )  }{ 2 - \sin^2(  \overline{ k_{so} }   L  ) }.
\end{align}
Note that since $\mathcal{C}_0$ or $\mathcal{E}_0$ in the denominator and the numerator cancel, the ratio does not depend on the energy, instead it depends only on the ratio between nanowire length and the averaged spin-orbit length.
The ratio is no greater than one, i.e., $r \leq 1$, and it reaches one (zero) when the nanowire length is half-integer (integer) of the spin-orbit length.
We therefore can extract the averaged spin-orbit coupling strength in a hybrid nanowire using the above formula.\\

\begin{figure}
\begin{center}
\includegraphics[width=0.9\linewidth]{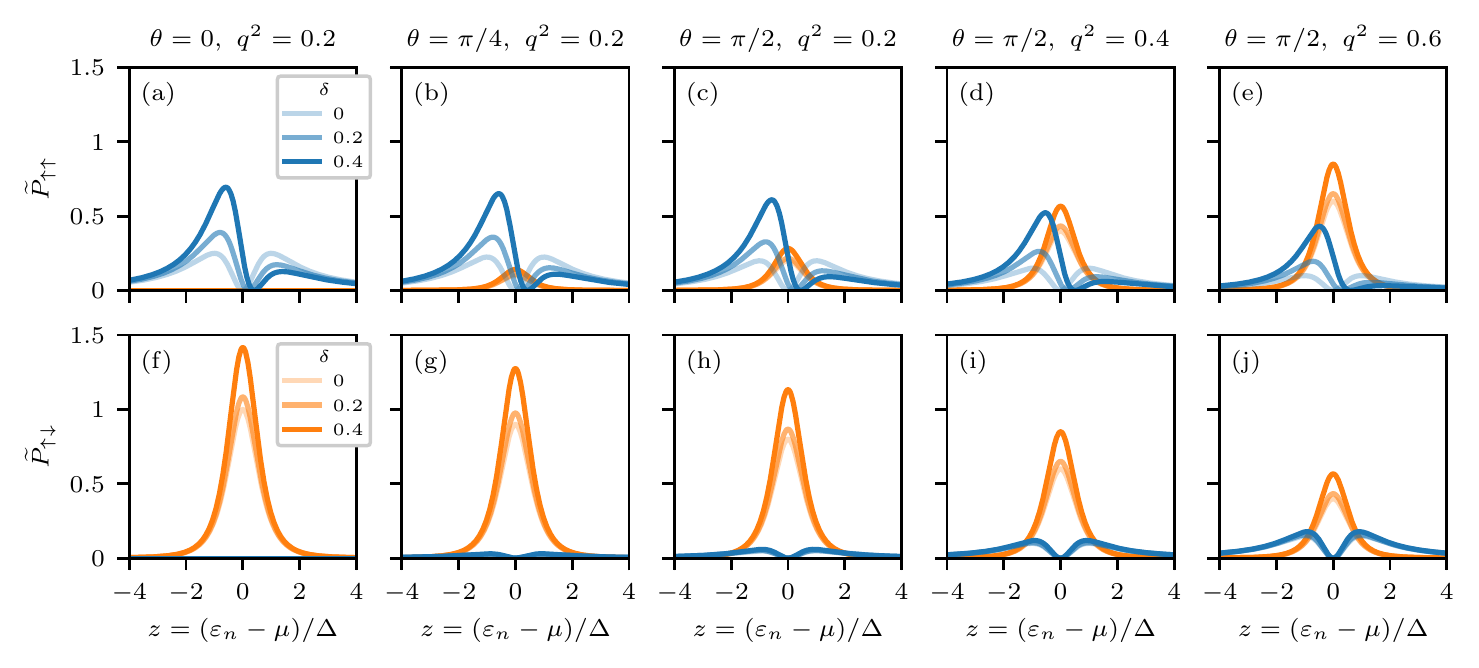}
\caption{We show CAR (orange) and ECT (blue) profiles at finite Zeeman splitting for various values of $\theta$ and $q^2$.
The response of these profiles to an increasing Zeeman field is identical to that in the main text.
}
\label{fig:SM_effect_Ez} 
\end{center}
\end{figure}

\subsection{$E_Z>0$}
We now consider a finite Zeeman field $E_Z\sigma_z$ inside the hybrid nanowire, with its direction being parallel to the dot spin axis.
It can be projected into the low-energy subspace spanned by the two normal states $\psi_n$ and $\psi_{\overline{n}}$:
\begin{align}
H_{Z,\mathrm{eff}} = E_Z
\begin{pmatrix}
\langle \psi_n |  \sigma_z | \psi_n  \rangle & \langle \psi_n |  \sigma_z | \psi_{\overline{n}}  \rangle \\
 \langle \psi_{\overline{n}} |  \sigma_z | \psi_n  \rangle & \langle \psi_{\overline{n}} |  \sigma_z | \psi_{\overline{n}} \rangle
\end{pmatrix},
\end{align}
with
\begin{align}
& \langle \psi_n |  \sigma_z | \psi_n  \rangle = -\langle \psi_{\overline{n}} |  \sigma_z | \psi_{\overline{n}} \rangle  = \int dx \phi^2(x) \left[ \cos^2(k_{so}x) + \sin^2(k_{so}x) \cos (2\theta) \right], \nn
& \langle \psi_n |  \sigma_z | \psi_{\overline{n}}  \rangle = \langle \psi_{\overline{n}} |  \sigma_z | \psi_n  \rangle^* =  \int dx \phi^2(x) \left[ -i \sin( \theta) \sin(2k_{so}x) + \sin^2(k_{so}x) \sin (2\theta) \right].
\end{align}
In the weak spin-orbit interaction $k_{so}L \ll 1$, the diagonal terms in are much larger than the off-diagonal ones, and thereby $H_{Z,\mathrm{eff}} \approx E_Z \sigma_z$ in the low-energy subspace.
We thus have the following low-energy effective Hamiltonian for a pair of Andreev bound states
\begin{align}
H_{\mathrm{eff}} = \xi_n ( a\dg_n a_n + a\dg_{\overline{n}} a_{\overline{n}} ) 
+ E_Z ( a\dg_n a_n - a\dg_{\overline{n}} a_{\overline{n}} ) 
+ \Delta( a\dg_n a\dg_{\overline{n}} + a_{\overline{n}} a_n ).
\end{align}
Since the Zeeman effect is diagonal in this projected basis, the eigenenergies and Bogoliubov wavefunctions for the Andreev bound states are
\begin{align}
&E_1 = \sqrt{\xi^2 + \Delta^2} + E_Z, \quad u_1(x \sigma) = u_0 \cdot \psi_n(x \sigma), \quad v_1(x \sigma) = v_0 \cdot \psi^*_{\overline{n}}(x \sigma), \nn
&E_2 = \sqrt{\xi^2 + \Delta^2} - E_Z, \quad u_2 (x \sigma) = - u_0 \cdot \psi_{\overline{n}}(x \sigma), \quad v_2 (x \sigma) = v_0 \cdot \psi^*_n(x \sigma).
\end{align}
Plugging them into Eq.~(3), we get
\begin{align}
& \wP^{\mathrm{CAR}}_{\su \su}(\delta) = \wP^{\mathrm{CAR}}_{\sd \sd}(\delta) = (z^2 + 1 - \delta^2)^{-2}  \cdot q^2  \sin^2 (\theta), \nn
& \wP^{\mathrm{CAR}}_{\su \sd}(\delta) = \wP^{\mathrm{CAR}}_{\sd \su}(\delta)  = (z^2 + 1 - \delta^2)^{-2}  \cdot \left[ p^2 + q^2 \cos^2 (\theta) \right], \nn
& \wP^{\mathrm{ECT}}_{\su \su}(\delta) =  \wP^{\mathrm{ECT}}_{\sd \sd}(-\delta)  =(z^2 + 1 - \delta^2)^{-2}  \cdot \left[  ( p z - \delta' )^2   +  q^2 \cos^2(\theta) (z - \delta)^2 \right], \nn
& \wP^{\mathrm{ECT}}_{\su \sd}(\delta)=\wP^{\mathrm{ECT}}_{\sd \su} (\delta)  =(z^2 + 1 - \delta^2)^{-2}  \cdot \left[  q^2 z^2 + \delta^2\cos^2(\theta)(1 - p)^2 \right]  \cdot   \sin^2(\theta),
\label{eq:P_Ez}
\end{align}
where $\delta = E_Z / \Delta$, and $\delta'=\delta( p \cos^2\theta +  \sin^2\theta )$.

In the main text, we show CAR and ECT profiles in Eq.~\eqref{eq:P_Ez} for increasing Zeeman spin splitting in the hybrid nanowire at  $\theta=\pi/2$ and $q^2=0.2$.
Now, we show more CAR and ECT profiles for different values of $\theta$ and $q^2$.
As shown in Fig.~\ref{fig:SM_effect_Ez}, the features of CAR and ECT profiles at finite $E_Z$ shown in the main text is general and appear for other choice of parameters as well.
That is, $\wP^{\rm{CAR}}_{\su \su}$, $\wP^{\rm{CAR}}_{\su \sd}$, and $\wP^{\rm{ECT}}_{\su \sd}$ all increase with $E_Z$, with their profiles remaining symmetric about $z=0$. 
By contrast, $\wP^{\rm{ECT}}_{\su \su}$ becomes asymmetric, with one peak being lifted and the other suppressed.
At the quantitative level, the profile amplitudes in the unfavorable channels, i.e, CAR-$\su \su$ and ECT-$\su \sd$ increase with $\theta$ and strength of spin orbit coupling characterized by $q^2=\sin^2(k_{so}L)$ in the range of $0<\theta<\pi/2$ and $q^2<1$.
By contrast, profile amplitudes in the favorable channels, i.e, CAR-$\su \sd$ and ECT-$\su \su$ decrease with $\theta$ and $q^2$.
At $\theta=0$ in panel (a) of Fig.~\ref{fig:SM_effect_Ez}, i.e., spin-orbit field and magnetic field are parallel, CAR-$\su \sd$ and ECT-$\su \su$ vanish completely due to spin conservation.

\section{Numerical simulations of realistic nanowires}

In this section, we numerically calculate CAR and ECT profiles for a semiconductor-superconductor nanowire.
The Bogoliubov-de Gennes Hamiltonian for the hybrid nanowire is
\begin{align}
H_{BdG} = \left( -\frac{\hbar^2}{2m^*}\partial^2_x - \mu  \right) \tau_z 
-i \alpha_R \partial_x \left(  \sin \theta \sigma_x + \cos \theta \sigma_z \right)
 + E_Z \tau_z \sigma_z + \Delta \tau_y \sigma_y.
\end{align}
Here, physical parameters are chosen according to Refs.~\cite{Wang2022Singlet}.
$m^*=0.015~m_e$ is the effective mass of InSb nanowire, $\mu$ is the chemical potential in the nanowire, $\alpha_R=0.12$ eV\angstrom~is the strength of the Rashba spin-orbit coupling, $\theta=\pi/2$ is the angle between Zeeman and spin-orbit field, $E_Z=\half \mu_B g B \approx  0.06$ meV is the induced Zeeman spin splitting, with $\mu_B$ the Bohr magneton, $g\approx 0.5  g_{InSb}\approx 20$ including the hybridization effect between InSb and Al, and the applied magnetic field is 0.1 T.
$\Delta = 0.2$ meV is the induced pairing potential which is slightly smaller than the parent Al gap. 
The length of the hybrid nanowire is $L=200$ nm.
For this nanowire, 
\begin{align}
k_{so}L \approx 0.47, \quad q^2 = \sin^2(k_{so}L) \approx 0.2,
\end{align}
putting it in the weak spin-orbit interaction regime.

\begin{figure}
\begin{center}
\includegraphics[width=0.9\linewidth]{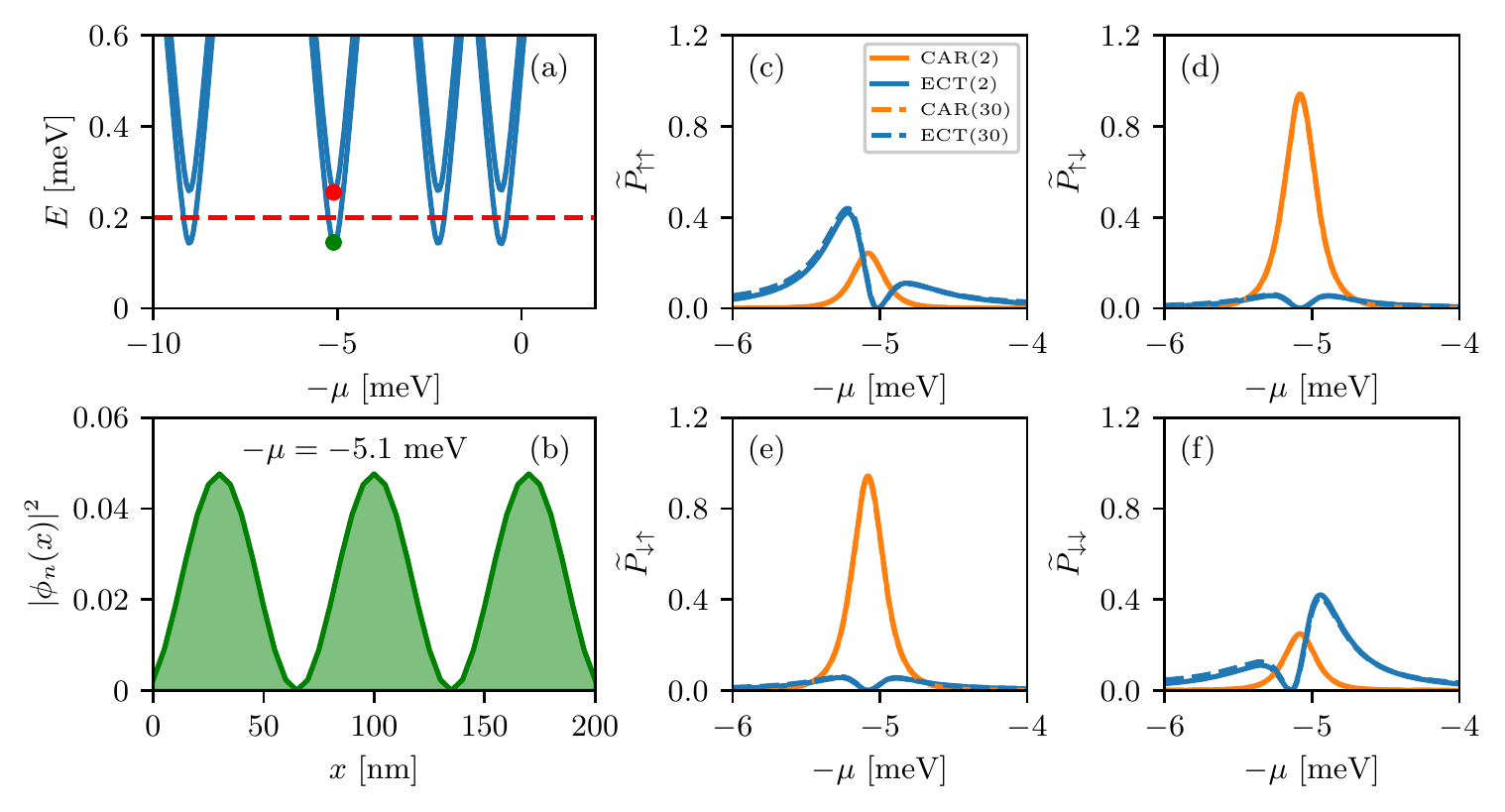}
\caption{Numerical simulations of a clean hybrid nanowire.
(a) Excitation energies as a function of chemical potential.
(b) Local density of states profile for the normal bound state at $\mu \approx 5.1$ meV.
(c)-(f) Solid lines: CAR and ECT profiles mediated by two Andreev bound states closest to the Fermi energy.
Dashed lines: CAR and ECT profiles mediated by thirty Andreev bound states.
Here $\phi^2_n(x_l) \phi^2_n(x_r) = 5.5 \times 10^{-6}$.
}
\label{fig:clean_wire} 
\end{center}
\end{figure}

We then discretize the above continuum Hamiltonian into a tight-binding Hamiltonian matrix with lattice constant $a=5$ nm using KWANT~\cite{kwant}.
The eigenenergies and eigenfunctions are obtained by diagonalizing the BdG Hamiltonian.
Importantly, we first calculate the bare $P^a$'s using Eq.~(4) in the main text.
In order to check whether a single pair of Andreev bound states is a good approximation, we compare results from summation over only two Andreev bound states as well as those from all Andreev bound states.  
Then, by dividing the bare $P^a$'s by local density of states $\phi^2_n(x_l)\phi^2_n(x_r)$ of a particular bound states which give the dominant contribution to $P^a$'s, we obtain the renormalized  $\wP^a$'s, such that a direct comparison is possible between the analytical results in Fig.~3 in the main text and the numerical results.
Numerically calculated $\wP^a$'s for a realistic nanowire show excellent agreement with the analytical results.

Figure~\ref{fig:clean_wire}(a) shows the excitation spectrum for a clean hybrid nanowire, and Fig.~\ref{fig:clean_wire}(b) shows the local density of states for the bound states at $-\mu \approx -5.1$ meV, with $\phi^2_n(x_l) \phi^2_n(x_r) = 5.5 \times 10^{-6}$.
This bound state denoted by the green dot and that by red dot give the dominant contribution to the CAR and ECT within the range of $-6< -\mu < -4$.
Here, we intentionally choose $-\mu$ to be the $x$-axis variable, consistent with $z=(\varepsilon_n - \mu) / \Delta$ in the main text.
The solid lines in Figs.~\ref{fig:clean_wire}(c)-\ref{fig:clean_wire}(f) show the calculated CAR and ECT profiles from only two Andreev bound states above the Fermi energy, i.e., denoted by the green and red dots in Fig.~\ref{fig:clean_wire}(a).  
The results agree well with the analytic ones in the main text at both qualitative and quantitative levels.
In addition, the dashed lines show the CAR and ECT profiles from thirty Andreev bound states above the Fermi energy.
Here we choose $N=30$ because $E_{30}$ is close to hopping amplitude $t=\hbar^2/(2m^*a^2)$, which is a natural cutoff energy in the tight-binding model.
Plus, the summation already converges well for $N$ close to 30.
The deviation of ECT is small, and that of CAR is negligible, because CAR decays faster at large $|z|$ than ECT. 
The results indicate that a pair of Andreev bound states is a good approximation for describing CAR and ECT mediated by short nanowires where level spacing is large.

\begin{figure}
\begin{center}
\includegraphics[width=0.9\linewidth]{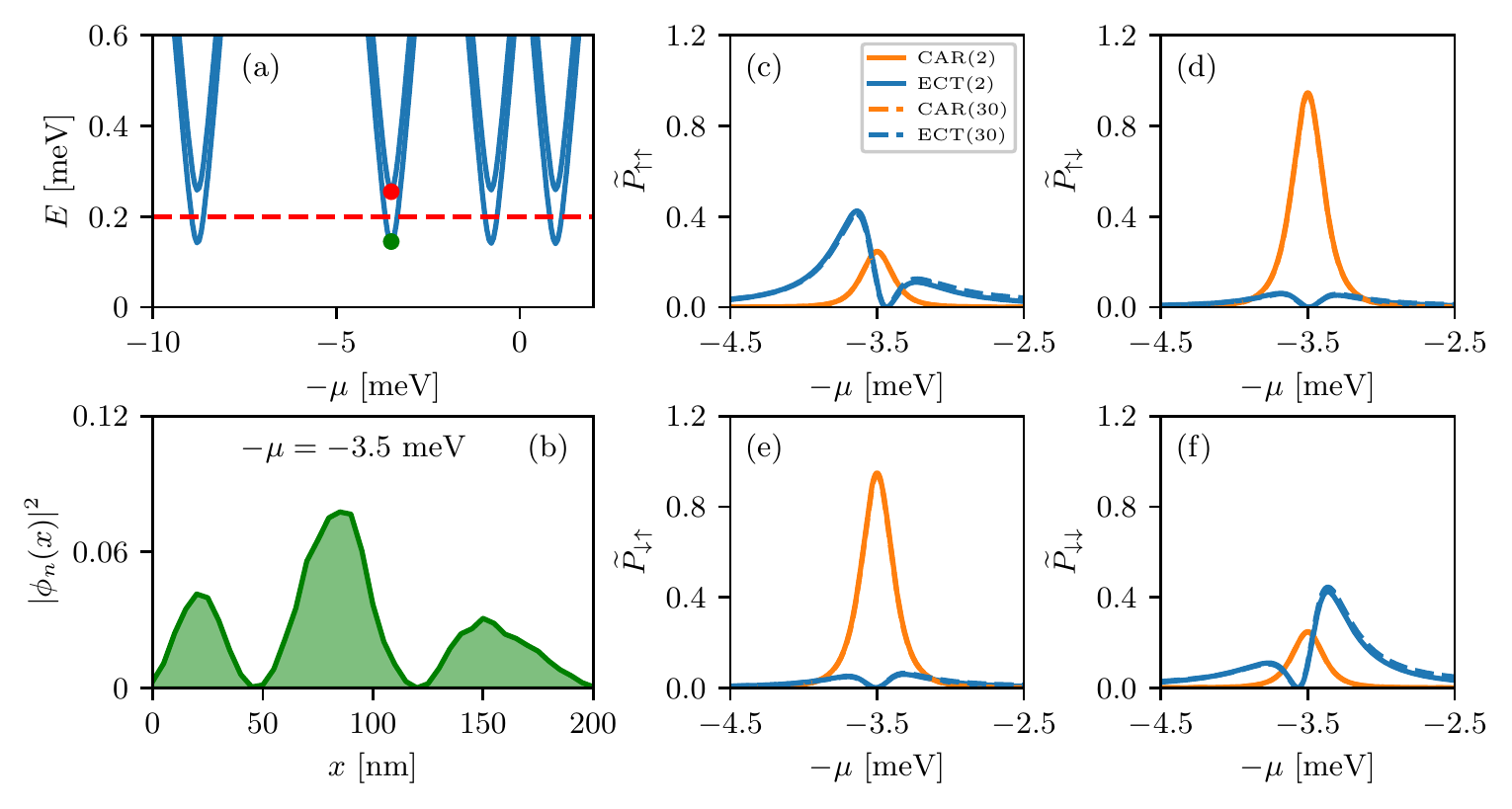}
\caption{Numerical simulations of a disordered hybrid nanowire with $U_D=10$ meV.
(a) Excitation energies as a function of chemical potential.
(b) Local density of states profile for the normal bound state at $\mu \approx 3.5$ meV.
(c)-(f) Solid lines: CAR and ECT profiles mediated by two Andreev bound states closest to the Fermi energy.
Dashed lines: CAR and ECT profiles mediated by thirty Andreev bound states.
Here $\phi^2_n(x_l) \phi^2_n(x_r) = 1.5 \times 10^{-6}$.
}
\label{fig:disorder_wire} 
\end{center}
\end{figure}

\begin{figure}
\begin{center}
\includegraphics[width=0.9\linewidth]{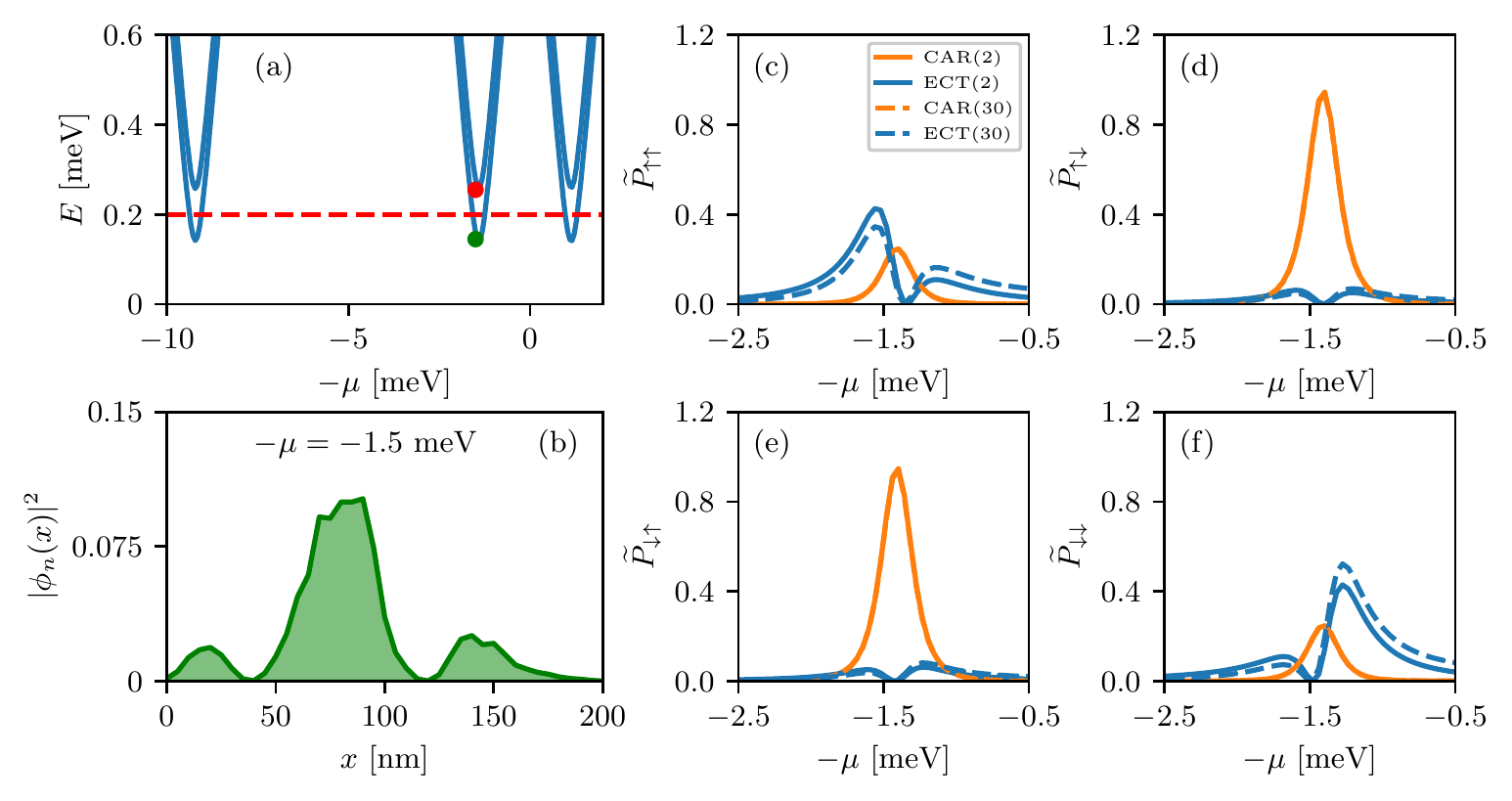}
\caption{Numerical simulations of a disordered hybrid nanowire with $U_D=20$ meV.
(a) Excitation energies as a function of chemical potential.
(b) Local density of states profile for the normal bound state at $\mu \approx 1.5$ meV.
(c)-(f) Solid lines: CAR and ECT profiles mediated by two Andreev bound states closest to the Fermi energy.
Dashed lines: CAR and ECT profiles mediated by thirty Andreev bound states.
Here $\phi^2_n(x_l) \phi^2_n(x_r) = 1.2 \times 10^{-7}$.
}
\label{fig:disorder_wire2} 
\end{center}
\end{figure}

In order to show that $\wP$'s depend only on the general features of the Andreev bound states, we also consider a disordered hybrid nanowire.
In particular, the effect of disorder is in the fluctuations of the chemical potential, i.e.,
\begin{align}
&\mu \to \mu + \delta \mu(x), \quad \delta \mu(x) \in [-\mu_D, \mu_D],
\end{align}
here we choose $U_D=10$ meV and $\delta \mu$ is independent for each lattice site.
The results are shown in Fig.~\ref{fig:disorder_wire}.
We see that now the Fermi energy of the bound state shifts from $-\mu=-5.1$ meV in a clean nanowire to $-3.5$ meV in the disordered one in Fig.~\ref{fig:disorder_wire}(a), and that the wavefunction profile becomes irregular Fig.~\ref{fig:disorder_wire}(b) and the corresponding local density of states is $\phi^2_n(x_l) \phi^2_n(x_r) = 1.5 \times 10^{-6}$, which is about four times smaller than the clean one.
Furthermore, we consider a nanowire subject to an even more disordered chemical potential, i.e., $U_D=20$ meV, as shown in Fig.~\ref{fig:disorder_wire2}.
Regardless of these changes due to disordered chemical potential, the profiles of $\wP$'s has negligible difference from those of the clean nanowire.
As explained in the main text, after the details of the orbital wavefunction being absorbed in the dot-hybrid coupling $t_{ln}$ and $t_{rn}$, the renormalized $\wP$'s relies only on the general properties of Andreev bound states, and thus its behavior is universal.

Thus, our numerical simulations show that the analytical results in the main text is applicable to realistic nanowire devices, and predicts general features of Andreev bound states.

\section{Resonant currents}

In this section, we give details of how we calculate the resonant current in the normal-dot-superconductor-dot-normal junction.
The methods we use include the $T$-matrix approach and the rate equation, which are standard for resonant current calculations~\cite{Recher2001Andreev, Loss2000Probing} in such mesoscopic systems. 
In terms of the parameter regime for generating resonant current, we consider 
\begin{align}
\Delta, U > \delta \mu > \Gamma_{DL}, k_B T, \quad \Gamma_{DL} > \Gamma_{SD}, \quad \varepsilon_l \approx \varepsilon_r \approx \mu_S,
\end{align}
which applies to both crossed Andreev reflection and elastic co-tunneling processes.

\subsection{Crossed Andreev reflection}
We first consider the CAR current, where two electrons pass from the superconductor via the virtual dot states to two different leads.
The whole tunneling process can be decomposed into two main parts.
In the first part, a Cooper pair breaks up, where one electron tunnels from superconductor to one dot level, leaving behind a quasiparticle excitation $E_m > \Delta$ in superconductor.
Almost simultaneously the second electron of the Cooper pair tunnels from superconductor to the other dot level before the first electron escapes from the dot to the electrode, because the relevant time scale is $\hbar/\Delta < \hbar/\Gamma_{DL}$.
Tunneling back to the superconductor is unlikely because $\Gamma_{SD} < \Gamma_{DL}$.
The amplitude for the transition from the initial to the final state is thus
\begin{align}
\langle f  | T(\varepsilon_i)  | i  \rangle \approx \langle f  | T(0)  | i  \rangle = \langle f  | T_2  | DD  \rangle \langle DD  | T_1  | i  \rangle,
\end{align}
where $T(0) = T(\varepsilon_i=0)$.
The initial state is $ | i \rangle = | 0_S \rangle | 0_D \rangle | \mu_l \rangle$, where the superconducting is in its ground state with no quasiparticle excitations, both dots levels are vacant and the normal leads are Fermi liquids filled up to its chemical potential.
$| DD \rangle =| 0_S \rangle | 1_l, 1_r \rangle | \mu_l \rangle $ is the intermediate state with dot states being occupied by one electron each.
$T_1$ is the $T$-matrix for the tunneling process in the first part is of second order in $H_{SD}$, with
\begin{align}
T_1 =  \frac{1}{i \eta - H_0} H_{SD}  \frac{1}{i \eta - H_0} H_{SD}.
\end{align}
Thus using the second-order perturbation theory we immediately have
\begin{align}
&\langle DD  | T_1  | i  \rangle = \frac{1}{ i \eta - ( \varepsilon_1 + \varepsilon_2 )} \cdot \Gamma^{\mathrm{CAR}}_{\eta \sigma}, 
\end{align}
with 
\begin{align}
 \Gamma^{\mathrm{CAR}}_{\eta \sigma} =  \frac{t_l t_r}{ \Delta} \sum_{m=1,2} \frac{ u_{m }(l \eta) v^*_{m }(r \sigma) - u_{m}(r\sigma) v^*_{m}(l\eta) }{E_m/\Delta}.
\end{align}
Here the spin in the dots are polarized in the $\sigma_z$ direction.
Since $\varepsilon_l \approx \varepsilon_r \approx \mu_S = 0$, the energy denominator diverges as $1/\eta$, indicating that the tunneling between dots and leads is resonant.
Thus for the second part of the tunneling process, we must include the tunnel Hamiltonian to all orders, i.e.,
\begin{align}
T_2 = H_{DL} \sum^{\infty}_{n=0} \left(  \frac{1}{i \eta - H_0} H_{DL} \right)^{2n+1},
\end{align}
and thus the transition amplitude for the second part is
\begin{align}
\langle f  | T_2  | DD  \rangle &= \Bigg \{ \left \langle pq \left |  H_{DL1} \right| Dq  \right \rangle \left \langle Dq \left | \sum^{\infty}_{m=0} \left( \frac{1}{i \eta - H_0} H_{DL1} \right)^{2m}  \right | Dq \right \rangle  \left \langle Dq \left | \frac{1}{i \eta - H_0} H_{DL2} \right| DD  \right \rangle \nn
&+ \left \langle pq \left |  H_{DL2} \right| pD  \right \rangle \left \langle pD \left | \sum^{\infty}_{m=0} \left( \frac{1}{i \eta - H_0} H_{DL2} \right)^{2m}  \right | pD \right \rangle  \left \langle pD \left | \frac{1}{i \eta - H_0} H_{DL1} \right| DD  \right \rangle \Bigg \} \nn
&\times \left \langle DD \left | \sum^{\infty}_{n=0} \left( \frac{1}{i \eta - H_0} H_{DL} \right)^{2n}  \right | DD \right \rangle.
\end{align}
Among all the above terms, the geometrical summation are calculated in a similar manner, and here we only focus on $\left \langle DD \left | \sum^{\infty}_{n=0} \left( \frac{1}{i \eta - H_0} H_{DL} \right)^{2n}  \right | DD \right \rangle$, which is
\begin{align}
\left \langle DD \left | \sum^{\infty}_{n=0} \left( \frac{1}{i \eta - H_0} H_{DL} \right)^{2n}  \right | DD \right \rangle
&\approx 1 + \sum^{\infty}_{n=1} \left( \left \langle DD \left | \left( \frac{1}{i \eta - H_0} H_{DL} \right)^2  \right | DD \right \rangle \right)^n \nn
&= \frac{1}{1 - \left \langle DD \left | \left( \frac{1}{i \eta - H_0} H_{DL} \right)^2  \right | DD \right \rangle },
\end{align}
with
\begin{align}
\left \langle DD \left | \left( \frac{1}{i \eta - H_0} H_{DL} \right)^2  \right | DD \right \rangle
&= \frac{1}{i\eta - (\varepsilon_1 +\varepsilon_2 )} \left( \frac{\Gamma_{DL2}}{2\pi} \int d \varepsilon_k \frac{1}{i \eta - (\varepsilon_1 + \varepsilon_k)} + \frac{\Gamma_{DL1}}{2\pi} \int d \varepsilon_k \frac{1}{i \eta - (\varepsilon_2 + \varepsilon_k)}  \right) \nn
&=\frac{1}{ \varepsilon_1 +\varepsilon_2 -i\eta} \left( \mathrm{Re} \Sigma + i \Gamma_{DL} \right),
\end{align}
where $\mathrm{Re} \Sigma =  \ln (E_{c1}/E_{c2} ) $ is a logarithmic divergence with the cutoff energy in the conduction band in the normal lead, and $\Gamma_{DL}=\Gamma_{DLl}+\Gamma_{DLr}$.
Note that here we assume that the dominant process is $| DD \rangle \to | pD \rangle \to | DD \rangle$ or $| DD \rangle \to | Dq \rangle \to | DD \rangle$, with $| pD \rangle$ denoting that the left electron is in the lead and the right electron in the dot.
So the geometrical summation is
\begin{align}
\left \langle DD \left | \sum^{\infty}_{n=0} \left( \frac{1}{i \eta - H_0} H_{DL} \right)^{2n}  \right | DD \right \rangle
= \frac{\varepsilon_1 +\varepsilon_2 -i\eta}{\varepsilon_1 +\varepsilon_2 -i \Gamma_{DL} }.
\end{align}
The key finding here is that the divergent denominator now becomes the numerator and the new denominator has a finite imaginary part which is proportional to the dot-lead coupling.
Finally the transition amplitude for the second part is
\begin{align}
\langle f  | T_2  | DD  \rangle &= \left( t'_l \frac{\varepsilon_1 +\varepsilon_q -i\eta}{\varepsilon_1 +\varepsilon_q -i \Gamma_{DLl} } \frac{t'_r}{i\eta - (\varepsilon_1 +\varepsilon_q)} +  t'_r \frac{\varepsilon_2 +\varepsilon_p -i\eta}{\varepsilon_2 +\varepsilon_p -i \Gamma_{DLr} } \frac{t'_l}{i\eta - (\varepsilon_1 +\varepsilon_q)} \right)  \frac{\varepsilon_1 +\varepsilon_2 -i\eta}{\varepsilon_1 +\varepsilon_2 -i \Gamma_{DL} } \nn
&= t'_l  t'_r\frac{\varepsilon_1 +\varepsilon_2 -i\eta}{ (\varepsilon_1 +\varepsilon_q -i \Gamma_{DLl})(\varepsilon_2 +\varepsilon_p -i \Gamma_{DLr}) }.
\end{align}
Plugging it into the formula for CAR current, we have
\begin{align}
I_{\mathrm{CAR}} &= \frac{e}{\hbar} \nu_1 \int d \varepsilon_p  \nu_2 \int  d \varepsilon_q | \langle pq  | T(0)  | i \rangle |^2 \delta(\varepsilon_p+\varepsilon_q ) \nn
&= \frac{e}{\hbar} \Gamma_{DLl} \Gamma_{DLr} |\Gamma^{\mathrm{CAR}}_{\eta \sigma}|^2 \int  d \varepsilon_p \frac{1}{ [( \varepsilon_1 -\varepsilon_p )^2 + \Gamma^2_{DLl} ]  [( \varepsilon_2 +\varepsilon_p )^2 + \Gamma^2_{DLr} ] } \nn
&= \frac{e}{\hbar} \cdot \frac{  \Gamma_{DL}  }{(\varepsilon_1+\varepsilon_2)^2 + \Gamma^2_{DL} } \cdot |\Gamma^{\mathrm{CAR}}_{\eta \sigma}|^2 .
\end{align}
So the CAR current has the form of a Breit-Wigner resonance profile, which assumes its maximum value at $\varepsilon_1+\varepsilon_2=0$.

\subsection{Elastic co-tunneling}
For the ECT current, a single electron passes from the lead with higher chemical potential via the dot and superconductor states to the other lead with lower chemical potential.
Here we make three assumptions in our derivation. First, when calculating the transition rate $W_{fi}$ between particular initial and final state, we assume that both normal leads are vacant. 
Second, the Fermi-Dirac distribution will be taken into account only in the final step for $\rho_{i}$. 
Third, when the chemical potential in leads are equal, the current flowing in the opposite directions cancel with each other. 
Under these assumptions, we first calculate the transition rate, focusing on the scenario of a single electron passing from the right lead to the left lead.
The total tunneling process can be separated into three parts.
\begin{align}
 \langle p | T(\varepsilon_i = \varepsilon_q) | q \rangle =  \langle p | T_3 | 1 \rangle \langle 1| T_2 | 2 \rangle \langle 2| T_1 | q \rangle,
\end{align}
where $ | l \rangle$ or $ | r \rangle$ means an electron is in the left or right dot.
For the first step, we need to include the resonant tunneling between dot and lead, such that
\begin{align}
\langle r| T_1 | q \rangle &= \left \langle r \left | \sum^{\infty}_{n=0} \left( \frac{1}{i\eta - H_0} H_{DLr} \right)^{2n} \right | r \right \rangle \left \langle r \left | \frac{1}{i\eta - H_0} H_{DLr} \right | q \right \rangle \nn
&= \frac{ t'_{r} }{\varepsilon_r - \varepsilon_q - i \Gamma_{DLr}}.
\end{align}
And 
\begin{align}
\langle l | T_2 | r \rangle &= \left \langle l \left | \left( \frac{1}{i\eta - H_0} H_{SD} \right)^2 \right | r \right \rangle \nn
&= \frac{ \Gamma^{\mathrm{ECT}}_{\eta \sigma} }{i \eta - ( \varepsilon_l - \varepsilon_q )},
\end{align}
with 
\begin{align}
\Gamma^{\mathrm{ECT}}_{\eta \sigma}=\frac{t_lt_r}{\Delta}  \sum_{m=1,2} \frac{ u_m(l \eta)  u^*_m(r \sigma) - v_m(r \sigma) v^*_m(l \eta) }{E_m/\Delta}.
\end{align}

For the third step, we have
\begin{align}
\langle p | T_3 | 1 \rangle &= \langle p | H_{DLl} | l \rangle \left \langle l \left | \sum^{\infty}_{n=0} \left( \frac{1}{i\eta - H_0} H_{DLl} \right)^{2n}  \right | l \right \rangle \nn
& = t'_l \frac{\varepsilon_l - \varepsilon_q - i \eta}{ \varepsilon_l - \varepsilon_q - i\Gamma_{DLl} }.
\end{align}
Therefore we have the transition amplitude and the transition rate to be
\begin{align}
\langle p | T | q \rangle &= \frac{ t'_l t'_r \Gamma^{\mathrm{ECT}}_{\eta \sigma}  }{( \varepsilon_l - \varepsilon_q - i \Gamma_{DLl} )( \varepsilon_r - \varepsilon_q - i \Gamma_{DLr} )}, \nn
W_{pq} &= 2\pi | \langle p | T | q \rangle |^2 \delta( \varepsilon_p - \varepsilon_q ).
\end{align}
The ECT current now is
\begin{align}
I &= \frac{e}{\hbar} \sum_{f, i} W_{fi} \rho_i \nn
&= \frac{e}{\hbar} \nu_l \int d \varepsilon_p \nu_r \int^{\delta \mu/2}_{-\delta \mu/2} d \varepsilon_q  | \langle p | T | q \rangle |^2 \delta( \varepsilon_p-\varepsilon_q ) \nn
&= \frac{e}{\hbar} \int^{\delta \mu/2}_{-\delta \mu/2} d \varepsilon_q \frac{ \gamma_{DL1} \gamma_{DL2} | \Gamma^{\mathrm{ECT}}_{\eta \sigma} |^2  }{[ (\varepsilon_l - \varepsilon_q)^2 + \gamma^2_{DL1}/4 ][ (\varepsilon_r - \varepsilon_q)^2 + \Gamma^2_{DLr} ]}\nn
&= \frac{e}{\hbar} \cdot \frac{ \Gamma_{DL}  }{ (\varepsilon_l - \varepsilon_r)^2 + \Gamma^2_{DL} } \cdot  |\Gamma^{\mathrm{ECT}}_{\eta \sigma}|^2.
\end{align}
Note that the integral of the outgoing electron energy $\varepsilon_p$ disappear because of the energy conservation.
The integral window of the incoming electron energy $\varepsilon_q$ is because when $\delta \mu=0$ the system is in equilibrium and left-flowing and right-flowing currents cancel and $I=0$. 
So the net current is due to the window of the biased voltage.
The ECT current is also in the form of the Breit-Wigner resonance, and assumes its maximum value at $\varepsilon_l =\varepsilon_r$.

\end{document}